\documentclass[manuscript,screen]{acmart}
\AtBeginDocument{%
  }

\usepackage{caption}
\usepackage{subcaption}
\usepackage{soul}
\usepackage{comment}
\usepackage{enumitem}
\usepackage{algorithm}
\usepackage{algpseudocode}
\usepackage{amsmath}
\usepackage{amsfonts}

\setcopyright{acmlicensed}
\copyrightyear{2018}
\acmYear{2018}
\acmDOI{XXXXXXX.XXXXXXX}

\usepackage[acronym,toc,shortcuts]{glossaries}

\makeglossaries
\newacronym{DNS}{DNS}{Domain Name System}
\newacronym{DNSSEC}{DNSSEC}{Domain Name System Security Extensions}
\newacronym{DS}{DS}{Digital Signature}
\newacronym{IP}{IP}{Internet Protocol}
\newacronym{KEM}{KEM}{Key Encapsulation Mechanism}
\newacronym{NIST}{NIST}{National Institute of Standards and Technology}
\newacronym{PQC}{PQC}{Post-Quantum Cryptography}
\newacronym{3GPP}{3GPP}{The 3rd Generation Partnership Project }
\newacronym{DDoS}{DDoS}{Distributed Denial of Service}
\newacronym{DoPQT}{DoPQT}{DNS over Post-Quantum TLS}
\newacronym{DoT}{DoT}{DNS over TLS}
\newacronym{DoH}{DoH}{DNS over HTTPS}

\begin{document}

\title{Quantum-Resistant Domain Name System: A Comprehensive System-Level Study}


\author{Juyoul Lee}
\affiliation{%
  \institution{Florida Institute of Technology}
  \city{Melbourne}
  \state{FL}
  \country{USA}}
\email{juyoul2023@my.fit.edu}

\author{Sanzida Hoque}
\affiliation{%
  \institution{Florida Institute of Technology}
  \city{Melbourne}
  \state{FL}
  \country{USA}}
\email{shoque2023@my.fit.edu}

\author{Abdullah Aydeger}
\affiliation{%
  \institution{Florida Institute of Technology}
  \city{Melbourne}
  \state{FL}
  \country{USA}}
\email{aaydeger@fit.edu}

\author{Engin Zeydan}
\affiliation{%
  \institution{Centre Tecnològic de Telecomunicacions de Catalunya (CTTC)}
  \city{Barcelona}
  \country{Spain}}
\email{engin.zeydan@cttc.cat}

\renewcommand{\shortauthors}{Lee et al.}

\begin{abstract}

The Domain Name System (DNS) plays a foundational role in Internet infrastructure, yet its core protocols remain vulnerable to compromise by quantum adversaries. As cryptographically relevant quantum computers become a realistic threat, ensuring DNS confidentiality, authenticity, and integrity in the post-quantum era is imperative. In this paper, we present a comprehensive system-level study of post-quantum DNS security across three widely deployed mechanisms: DNSSEC, DNS-over-TLS (DoT), and DNS-over-HTTPS (DoH). We propose \textit{Post-Quantum Cryptographic (PQC)-DNS}, a unified framework for benchmarking DNS security under legacy, post-quantum, and hybrid cryptographic configurations. Our implementation leverages the Open Quantum Safe (OQS) libraries and integrates lattice- and hash-based primitives into BIND9 and TLS 1.3 stacks. We formalize performance and threat models and analyze the impact of post-quantum key encapsulation and digital signatures on end-to-end DNS resolution. Experimental results on a containerized testbed reveal that lattice-based primitives such as Module-Lattice-Based Key-Encapsulation Mechanism (MLKEM) and Falcon offer practical latency and resource profiles, while hash-based schemes like SPHINCS+ significantly increase message sizes and processing overhead. We also examine security implications including downgrade risks, fragmentation vulnerabilities, and susceptibility to denial-of-service amplification. Our findings inform practical guidance for deploying quantum-resilient DNS and contribute to the broader effort of securing core Internet protocols for the post-quantum future.
\end{abstract}




\maketitle


\section{Introduction}

The \ac{DNS} is a fundamental basis of the Internet that translates human-readable domain names into \ac{IP} addresses and facilitates all modern network communication. However, the original DNS protocol was not designed with security in mind, leaving it vulnerable to a number of attacks. Common threats include \ac{DNS} spoofing (or cache poisoning), where attackers inject false DNS records into a resolver's cache, DNS amplification attacks, a form of \ac{DDoS}, where DNS servers are exploited to flood a target with traffic, and DNS hijacking, where queries are redirected to malicious servers \cite{Hudaib2014DNS}. Over the past two decades, a number of \ac{DNS} security extensions and enhancements, \ac{DNSSEC}, \ac{DoT}, and \ac{DoH}, have been developed to mitigate threats such as spoofing, tampering and surveillance \cite{10537516} \cite{10486930}. DNSSEC provides data integrity and origin authentication through digital signatures and thus protects against cache poisoning and similar attacks. DoT, and DoH, provide encrypted transport mechanisms to protect DNS requests and responses from eavesdropping and data manipulation in transit. Together, these protocols form the backbone of secure DNS resolution in the current Internet ecosystem.

However, the security of all three protocols depends on classical public key cryptographic primitives, most commonly RSA, ECDSA, and X25519, which are known to be vulnerable to quantum attacks. Once quantum computers can be deployed on a large scale, algorithms like Shor’s will render these cryptosystems obsolete and make DNS vulnerable to privacy and integrity issues \cite{aydeger2024towards}.

To counter this new threat, there is growing interest in \ac{PQC} - cryptographic algorithms designed to resist attacks from quantum computers. Although standardization efforts are underway, most notably by \ac{NIST}, adapting these new primitives to the DNS infrastructure is a major challenge. The DNS is a performance-critical system with strict constraints on message size, latency, and compatibility. Post-quantum signature schemes such as MLDSA, Falcon, and SPHINCS+, while secure against quantum adversaries, may introduce larger keys and signatures or increased computational overhead, which can impact DNSSEC's ability to remain efficient and interoperable. Similarly, post-quantum key exchange mechanisms must be evaluated in the context of DoT and DoH, where low-latency TLS handshakes are essential for practical use.

While progress has been made in incorporating post-quantum cryptographic primitives into particular components of DNS security, such as DNSSEC with PQ-signatures \cite{mcgowan2025security} or TLS with PQ key exchanges \cite{bozhko2023performance}, most current research endeavors focus on these protocols alone. This disjointed picture leads to an inadequate comprehension of the performance of post-quantum measures across the whole DNS security stack, especially when deployed in combination within real-world operational constraints. In particular, encrypted \ac{DNS} protocols such as \ac{DoT} and \ac{DoH}, which form the privacy-enhancing basis of current DNS resolution, have received little attention in the post-quantum discourse compared to \ac{DNSSEC}. Furthermore, the impact of the introduction of post-quantum cryptography on the interoperability, latency and scalability of DNS systems has hardly been comprehensively investigated when all three protocols are analysed together. This narrow focus creates a notable gap in the literature: the lack of a thorough, comparative, and system-level examination of DNSSEC, DoT, and DoH within post-quantum cryptography frameworks. Without this, it is difficult to assess the practical feasibility and trade-offs of transitioning DNS infrastructure to a post-quantum-resilient architecture.

To address this gap, our research presents a comprehensive analysis of post-quantum security concerning the three primary DNS protection mechanisms: DNSSEC, DoT, and DoH. We specifically present the following key contributions:

\begin{itemize}[leftmargin=*]
    \item \textbf{\textit{Post-Quantum DNS Implementation and Evaluation Framework:}}  
    We present a unified implementation of DNSSEC, DoT, and DoH protocols secured with NIST-recommended post-quantum cryptographic primitives. Our framework enables protocol-level benchmarking across classical, hybrid, and PQC-only modes, and includes a formal model of DNS resolution performance under varying cryptographic configurations. Experimental results quantify latency, bandwidth, and resource trade-offs, providing practical insights into DNS migration feasibility.

    \item \textbf{\textit{Security Threat Taxonomy and Mitigation Strategies:}}  
    We identify critical vulnerabilities introduced by PQC adoption-such as downgrade attacks, timing leaks, and fragmentation exploits-and propose mitigation strategies ranked by severity. Our analysis informs secure-by-design integration of PQC into DNS systems.

    \item \textbf{\textit{Deployment Challenges and Design Recommendations:}}  
    We uncover practical constraints-including compatibility gaps, resource consumption at the edge, and hybrid mode fragility-that limit immediate deployment. Based on our findings, we offer actionable guidance for transitioning DNS infrastructure toward quantum resilience.
\end{itemize}

The remainder of the paper is structured as follows. Section~\ref{sec:related} surveys related work on DNS security extensions and post-quantum cryptography in networked protocols. Section~\ref{sec:design} presents the system architecture and formalizes the performance model, protocol sequence, and associated cost equations for PQC-DNS. Section~\ref{sec:design} also analyzes emerging security threats introduced by PQC adoption, and presents corresponding mitigation strategies. Section~\ref{sec:results} reports empirical findings for DNSSEC, DoT, and DoH under classical, hybrid, and post-quantum cryptographic configurations. Section~\ref{sec:discussion} discusses deployment challenges, sustainability, and directions for future work. Finally, Section~\ref{sec:conclusion} concludes the paper.

\section{Related Work}
\label{sec:related}
DNS, a foundational component of the Internet, has long been recognized as a target of privacy and integrity attacks. In response, several privacy enhancements such as DoT~\cite{RFC7858}, DoH~\cite{RFC8484}, and DNSSEC~\cite{RFC4033} have been standardised to enable encryption and authentication. However, these protocols are based on classical cryptographic primitives that are vulnerable to future quantum adversaries. With the standardisation of PQC underway~\cite{NISTPQC}, recent research has begun exploring how DNS security protocols can be adapted for post-quantum resilience. Early work on PQ-TLS, e.g. by Stebila et al.~\cite{Stebila2016Hybrid} and Hülsing et al.~\cite{Huelsing2021PQTLS}, demonstrated hybrid key exchange techniques that combine classical and post-quantum algorithms (e.g. X25519+MLKEM) within the TLS 1.3 framework. These hybrid approaches have been piloted in real-world deployments, including Google Chrome~\cite{GooglePQ2023} and Cloudflare~\cite{CloudflarePQDNS}, and showed acceptable latency and bandwidth overheads. However, these evaluations primarily focused on performance metrics rather than broader integration into DNS infrastructures.

Work on incorporating post-quantum cryptography for DNS has been comparatively limited. Titan-DoH by Ali and Chen \cite{ali5230452titan} introduces a trust-aware, adaptive architecture for PQC-secure DoH. Their solution integrates trust algebra, graph signal processing, Bayesian contextual inference, and FrodoKEM-based TLS handshakes, supported by verifiable delay functions to identify encrypted malicious requests. The system demonstrates high accuracy and low latency under simulated adversarial loads, pushing the boundaries of secure PQC-ready DoH infrastructures.  

Other existing work focuses primarily on DNSSEC. The IETF DNSOP working group has explored considerations for integrating post-quantum signatures into DNSSEC, addressing issues like key size, algorithm agility, and protocol compatibility~\cite{IETFPQDNSSEC}. Zhang et al.~\cite{Zhang2022SPHINCS} evaluated SPHINCS+ in a DNSSEC setting, noting its suitability for quantum-resilient authentication but also highlighting challenges due to its large signature sizes and the risk of IP-layer fragmentation. These concerns underscore the importance of experimentally validating PQ signatures within real DNS server and resolver implementations.  Pan et al. \cite{pan2024double} present a double signature DNSSEC method that integrates classical and post-quantum techniques to provide transitional security against quantum and classical attacks. Through the use of application layer fragmentation, their dual-signed records provide resolution over UDP while adhering to packet size limitations. Raavi et al. \cite{raavi2024securing} tackle fragmentation-based attacks using a commit-and-reveal method with a blockchain-based public key offloading strategy. Their approach guarantees fragment authenticity and lowers DNSSEC packet size, especially for substantial signatures such as Falcon-512. The solution, though unique, addresses fragmentation mis-association issues inside DNSSEC and does not integrate with wider DNS security protocols like DoT and DoH. 

Goertzen and Stebila \cite{goertzen2023post} propose ARRF (Application-layer Request-based Resource Fragmentation), which moves fragmentation logic to the application layer. Unlike prior fragmentation methods, ARRF sends an initial truncated response and requires the client to request additional fragments explicitly, improving both reliability and backward compatibility. Their experiments show ARRF significantly reduces resolution time and data overhead compared to traditional DNS-over-UDP with TCP fallback when using PQC algorithms such as Falcon-512, Dilithium2 (aka MLDSA44), and SPHINCS+. Expanding upon ARRF, McGowan et al. \cite{mcgowan2025security} identify a memory exhaustion vulnerability that can be exploited by altered RRSIZE fields. They address this using a dynamic memory allocation approach that maintains ARRF’s efficiency while protecting against amplification threats. 
Rawat and Jhanwar have contributed several complementary protocols aimed at minimizing the overhead of PQC integration. They present QNAME-Based Fragmentation (QBF) \cite{rawat2023post}, a DNS-layer fragmentation scheme that avoids IP fragmentation and TCP fallback altogether. QBF fragments DNSSEC responses using standard DNS records, enabling reconstruction in a single round trip without altering the DNS protocol stack. Their experimental results demonstrate significant performance gains: QBF outperforms both standard DNS and parallel ARRF in post-quantum resolution scenarios, particularly with Falcon-512, Dilithium2 (aka MLDSA44), and SPHINCS+. By remaining fully backward compatible and avoiding changes to zone files or DNS stacks, QBF represents a promising middle ground for immediate PQC deployment. Rawat and Jhanwar further extend this line of work with the SL-DNSSEC protocol \cite{rawat2024quantum}, which replaces digital signatures with post-quantum \glspl{KEM} and MACs, significantly decreasing message size and resolution delay. TurboDNS \cite{rawat2024post} enhances PQC DNSSEC over TCP by including authentication data into the first UDP query, using cryptographic cookies to provide one-round-trip resolution. While these protocols offer substantial performance gains, they depart from traditional signature-based validation chains, posing challenges for universal adoption. In parallel, Schutijser et al. \cite{schutijser2024testbed} introduce PATAD, a containerized platform for PQC and DNSSEC experimentation. PATAD employs PowerDNS and modular topologies to assess Falcon and other signature methods inside realistic zone hierarchies. While PATAD enables empirical benchmarking of Falcon and other candidates, it remains focused on DNSSEC and does not extend to encrypted DNS protocols such as DNS-over-TLS (DoT) or DNS-over-HTTPS (DoH).

Several theses offer foundational empirical studies supporting the integration of post-quantum signatures in DNSSEC. Jafarli \cite{jafarli2022providing} presents a Merkle Tree-based framework using XMSS for the signing of grouped records. This framework optimizes payload dimensions and signing efficacy by adjusting tree size according to record update frequency, but at the expense of heightened memory consumption and update intricacy.
Beernink \cite{beernink2022taking} assesses the viability of many post-quantum cryptography algorithms using actual DNS traffic and signer logs. His research demonstrates that Falcon-512 complies with current DNS packet limitations while exerting little computing burden. Beernink further presents an out-of-band key exchange architecture to bypass signature size constraints for more substantial post-quantum cryptography systems such as Rainbow.
Projects like \textit{PQDNS}~\cite{PQDNS2022} and industry recommendations from NLnet Labs~\cite{NLnetLabsPQC} and NIST~\cite{NISTIR8105} have emphasized the importance of prototyping and testing post-quantum secure DNS under realistic operational conditions. However, most existing works focus on isolated components (e.g., PQ-TLS or PQ-DNSSEC) rather than a holistic evaluation across all major DNS security channels.

\section{Proposed PQC-DNS Method}
\label{sec:design}
In this section, we first give a short background on the DNS infrastructure and how the DNS protocol works. Later, we describe the problem that we aim to solve. Then, we define our proposed PQC-DNS sequences, and finally we discuss security threats and potential mitigation techniques.

\subsection{DNS Infrastructure}
\begin{figure}
    \centering
    \includegraphics[width=0.56\linewidth]{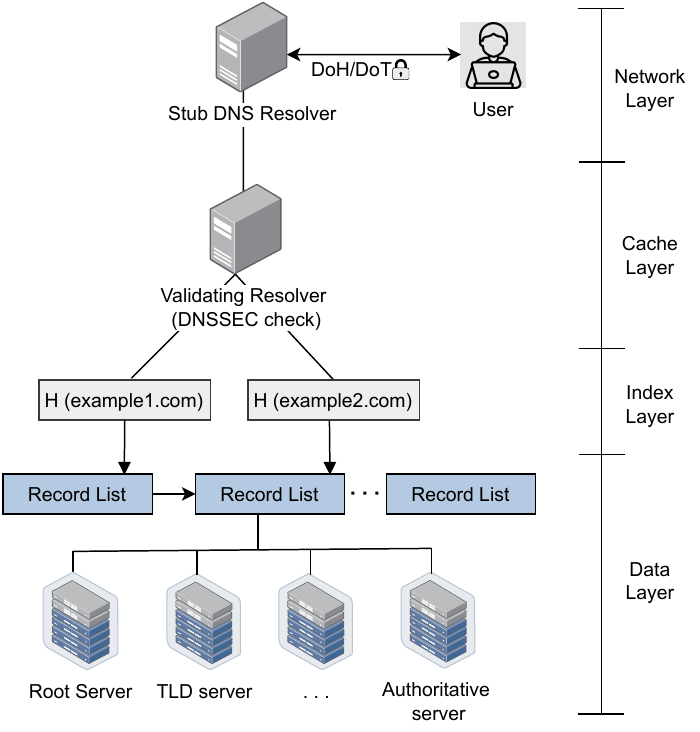}
    \caption{Layered Architecture of Secure \ac{DNS}}
    \label{fig:dns}
\end{figure}

Fig. \ref{fig:dns} presents a vertically layered architecture of secure \ac{DNS} resolution that integrates \ac{DoH}, \ac{DoT}, and \ac{DNSSEC} into a coherent workflow. It is structured across four key layers: Network, Cache, Index, and Data, each representing a distinct phase in the resolution and security pipeline. At the Network Layer, a user initiates a DNS query that is transmitted securely to a DNS resolver using encrypted transport protocols (\ac{DoH} or \ac{DoT}), ensuring confidentiality and protection against man-in-the-middle attacks. The resolver forwards the query to a Validating Resolver in the Cache Layer, which checks whether the result is cached and performs \ac{DNSSEC} validation by verifying digital signatures (RRSIG) against trusted keys (DNSKEY and DS). If the answer is not locally available, the resolver consults the Index Layer, where the queried domain (e.g., example.com) is hashed and mapped to a pointer that locates the relevant \ac{DNS} records. Finally, the Data Layer contains the actual signed \ac{DNS} records retrieved from external authoritative DNS servers, such as the Root, TLD (.com), and domain-specific authoritative name servers. These servers return the required resource records (A, AAAA, DNSKEY, RRSIG), which are then validated recursively up to the \ac{DNS} root using \ac{DNSSEC}. The validated result is passed back up through the layers and ultimately returned to the user. This layered model demonstrates how privacy (via \ac{DoH}/\ac{DoT}) and integrity (via \ac{DNSSEC}) are simultaneously enforced in modern DNS infrastructure.

\subsection{Problem Formulation}
\label{ss:prob}
While the security properties of \ac{PQC} algorithms have been rigorously analyzed in isolation, their systematic integration into DNS protocols remains underexplored, especially for encrypted transport mechanisms such as \ac{DoT} and \ac{DoH}. Modern transition frameworks, such as those proposed by NIST and the IETF, include both PQC-only and hybrid classical+PQC configurations to accommodate backward compatibility. However, their real-world impact on latency, resource consumption, and protocol compatibility requires a unified system-level evaluation. To address this, we propose a unified post-quantum DNS security architecture (PQC-DNS) that evaluates three deployment classes:

\begin{enumerate}
  \item \textbf{Legacy-only:} DNS configurations using classical \ac{KEM} and signature algorithms.
  \item \textbf{PQC-only:} DNS configurations using post-quantum \glspl{KEM} and signatures.
  \item \textbf{Hybrid:} Mixed deployments combining classical and \ac{PQC} schemes in KEM or digital signatures.
\end{enumerate}

We define a DNS resolution performance profile $\mathcal{P}_{\text{DNS}}^{(k, s)}$ as a function of cryptographic configurations, as described in Equation~\ref{eq:perf}. Notations used throughout the paper are listed in Table \ref{tab:notation}.

\begin{equation}\label{eq:perf}
\mathcal{P}_{\text{DNS}}^{(k, s)} = \delta_{\text{DNSSEC}} \cdot \mathcal{S}_{\text{DNSSEC}}(k, s) + \mathcal{S}_{\text{Transport}}(k, s)
\end{equation}

Each $\mathcal{S}_i(k, s)$ is defined as:
\begin{equation}
\mathcal{S}_i(k, s) = \left( T_{\text{latency}}, B_{\text{bandwidth}}, C_{\text{client}}, C_{\text{server}}, M_{\text{client}} \right)
\end{equation}

\begin{table}[htbp]
\centering
\caption{Notation Summary for PQC-DNS Performance Modeling}
\label{tab:notation}
\begin{tabular}{|p{3.2cm}|p{10cm}|}
\hline
\textbf{Symbol} & \textbf{Meaning} \\
\hline
$\mathcal{P}_{\text{DNS}}^{(k, s)}$ & DNS resolution performance profile under KEM $k$ and signature $s$ scheme \\
\hline
$k$ & Selected post-quantum Key Encapsulation Mechanism (e.g., MLKEM) \\
\hline
$s$ & Selected post-quantum Signature Scheme (e.g., MLDSA, Falcon, SPHINCS+) \\
\hline
$\delta_{\text{DNSSEC}}$ & Indicator variable for whether DNSSEC is enabled ($1$) or disabled ($0$) \\
\hline
$\mathcal{S}_i(k, s)$ & Measured performance metrics under $(k, s)$ for component $i$ \\
\hline
$T_{\text{latency}}$ & End-to-end resolution latency \\
\hline
$B_{\text{bandwidth}}$ & Total bandwidth consumed for DNS transaction \\
\hline
$C_{\text{client}}, C_{\text{server}}$ & CPU usage on client and server during resolution \\
\hline
$M_{\text{client}}$ & Memory overhead on the client \\
\hline
$T_{\text{Phase1}}$ & Time to establish a PQC-secure TLS 1.3 session \\
\hline
$T_{\text{CH}}, T_{\text{SH}}$ & Time to send/receive \texttt{ClientHello} / \texttt{ServerHello} \\
\hline
$T_{\text{KEM}}, T_{\text{SIG}}$ & Time for key encapsulation and signature verification \\
\hline
$T_{\text{KDF}}$ & Time for TLS key derivation function \\
\hline
$T_{\text{FIN}}$ & Time to exchange TLS \texttt{Finished} messages \\
\hline
$T_{\text{TLS\_termination}}$ & Time for TLS session termination or reset \\
\hline
$n$ & Number of DNS resolution steps (root, TLD, authoritative) \\
\hline
$T_{\text{query},i}, T_{\text{response},i}$ & Time to send query and receive response from DNS server $i$ \\
\hline
$T_{\text{DNSSEC},i}$ & Time to validate DNSSEC signature at server $i$ \\
\hline
$T_{\text{return}}$ & Time to return final validated DNS response to client \\
\hline
\end{tabular}
\end{table}

This formulation allows systematic comparison across cryptographic configurations and quantifies trade-offs in latency, computational effort, bandwidth, and memory.

\vspace{1em}
\noindent\textbf{Security Model:}
We consider an adversary $\mathcal{A}$ with quantum computational capabilities, capable of executing Shor’s and Grover’s algorithms. The attacker may (i) Eavesdrop and perform MITM attacks on TLS handshakes. (ii)  Perform DNS spoofing and cache poisoning. (iii) Attempt to break digital signatures or key exchanges via quantum means. We assume PQC primitives satisfy the following: (i) KEMs are IND-CCA2 (Indistinguishability of Ciphertexts under Chosen-Ciphertext Attack) secure under quantum adversaries \cite{chevalier2022security}. (ii) Signatures are EUF-CMA (Existential Unforgeability under Chosen Message Attack) secure in the quantum random oracle model \cite{xagawa2024signatures}. The system seeks to provide post-quantum confidentiality, authenticity, and integrity for DNS queries and responses against such adversaries.

\subsection{PQC-DNS Protocol Sequence}

\begin{figure}[htp!]
    \centering
    \includegraphics[width=0.85\linewidth]{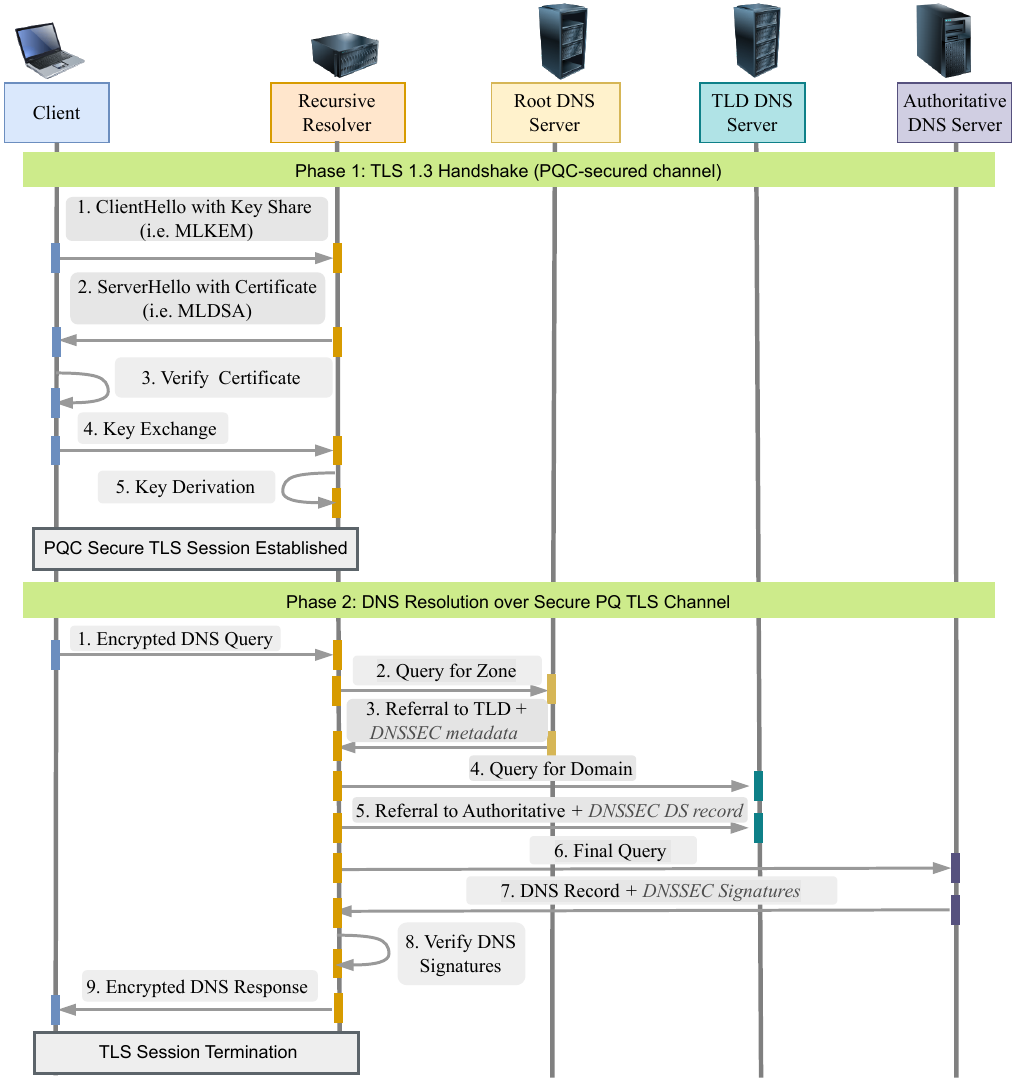}
    \caption{Layered Protocol Sequence for PQC-DNS with DNSSEC Option}
    \label{fig:dnspqc}
\end{figure}

Fig.~\ref{fig:dnspqc} illustrates the architecture of PQC-DNS. The protocol consists of two stages: a TLS 1.3 handshake with PQC primitives and the subsequent DNS resolution.

\vspace{1em}
\noindent\textbf{Phase 1: Establishing PQC-Secure TLS 1.3 Session}

Let $T_{\text{Phase1}}$ be the total time to establish a post-quantum secure TLS session. We define:

\begin{equation}
T_{\text{Phase1}} = T_{\text{CH}} + T_{\text{SH}} + T_{\text{KEM}} + T_{\text{SIG}} + T_{\text{KDF}} + T_{\text{FIN}} + T_{\text{TERM}}
\end{equation}

\noindent This handshake ensures post-quantum confidentiality and mutual authentication prior to DNS exchange.
\vspace{1em}

\noindent\textbf{Phase 2: Encrypted Recursive DNS Resolution}

Once TLS is established, DNS queries traverse the traditional recursive path under encryption. The total time for DNS query resolution is:

\begin{equation}
T_{\text{Phase2}} = \sum_{i=1}^{n} (T_{\text{query},i} + T_{\text{response},i}) + \delta_{\text{DNSSEC}} \cdot \sum_{i=1}^{n} T_{\text{DNSSEC},i} + T_{\text{return}}
\end{equation}

\noindent The variables appearing in these equations are defined in Table~\ref{tab:notation}.

\subsection{Security Considerations}

PQC-DNS resists quantum attacks due to the use of NIST-standardized algorithms. However, there are further threats to be considered as follows: \textit{(i) Timing Attacks:} Constant-time implementations of post-quantum KEM and signature algorithms are essential to prevent side-channel leakage during TLS handshakes. Recent work has shown that even lattice-based schemes like Kyber and Dilithium may be susceptible if implemented without strict timing controls~\cite{hochstatter2023leaky}. \textit{(ii) DDoS Attacks:} The increased bandwidth and CPU overhead introduced by PQC operations, especially during handshake and validation phases, can be exploited for denial-of-service attacks. Mitigation strategies include client puzzles and rate-limiting mechanisms tailored for PQC-induced latency~\cite{shulman2023quantumddos}. \textit{(iii) Downgrade Attacks:} Hybrid configurations combining classical and post-quantum algorithms must rigorously enforce cipher suite negotiation policies to prevent fallback to classical-only modes. Improper negotiation can expose sessions to downgrade attacks, negating the intended quantum resistance~\cite{ietf2023hybrid}. \textit{(iv) Fragmentation Vulnerabilities:} Large PQC signatures (e.g., SPHINCS+) may trigger IP-layer fragmentation, particularly in DNSSEC responses. Techniques such as \textit{Application-layer Request-based Resource Fragmentation (ARRF)}~\cite{goertzen2022arrf} mitigate this by moving fragmentation logic to the application layer. \textit{QNAME-Based Fragmentation (QBF)}~\cite{rawat2023qbf} avoids IP-layer fragmentation entirely by using standard DNS record fields. While protocols like \textit{TurboDNS}~\cite{rawat2023turbodns} and \textit{SL-DNSSEC}~\cite{rawat2023sldnssec} improve post-quantum DNSSEC performance, they do not directly address fragmentation and may pose compatibility trade-offs. \textit{(v) Key/Signature Reuse:} Reusing nonces or ephemeral keys in PQC schemes (e.g., Dilithium, Falcon) can lead to private key recovery via algebraic attacks. This risk is amplified in multithreaded resolver environments with shared randomness. Mitigation includes using hardened libraries (e.g., liboqs) that enforce one-time key usage and thread-safe randomness~\cite{dukhovni2023nonce}. \textit{(vi) Interoperability Failures:} PQC deployment may fail across clients, resolvers, or middleboxes lacking support for hybrid or PQC-only cipher suites. Misconfigured fallback behavior can result in silent downgrades or dropped connections. Operators should test compatibility using hybrid-aware configurations and monitor failure modes~\cite{ietf2023interop}. Finally, the summary of these threats and potential mitigation strategies are listed in Table \ref{tab:threats}.

\begin{table}[htp!]
\centering
\footnotesize
\caption{Summary of Threats and Mitigations in PQC-DNS Deployment}
\label{tab:threats}
\begin{tabular}{|p{1.7cm}|p{5.3cm}|p{5.7cm}|p{1cm}|}
\hline
\textbf{Threat} & \textbf{Description and Exploit Vector} & \textbf{Mitigation Strategy} & \textbf{Severity} \\
\hline \hline
\textbf{Timing Attacks} & Variation in handshake duration reveals secret-dependent operations in KEM or signatures. & Use constant-time implementations of all cryptographic primitives; avoid branching on secret values~\cite{hochstatter2023leaky}. & High \\
\hline
\textbf{DDoS Amplification} & PQC handshakes consume more CPU/memory, making resolvers vulnerable to resource exhaustion via spoofed requests. & Deploy client puzzles, adaptive rate-limiting, and TLS session resumption to mitigate load~\cite{shulman2023quantumddos}. & High \\
\hline
\textbf{Downgrade Attacks} & Adversary forces fallback to classical-only modes in hybrid configurations by interfering with cipher negotiation. & Enforce hybrid binding and strict cipher suite policies; validate negotiated modes~\cite{ietf2023hybrid}. & High \\
\hline
\textbf{Fragmentation Attacks} & Large PQC signatures cause IP-layer fragmentation, which can be exploited for DNS poisoning or evasion. & Use application-layer fragmentation techniques (e.g., ARRF~\cite{goertzen2022arrf}) or QNAME-based approaches~\cite{rawat2023qbf}; avoid UDP-only transport. & Medium \\
\hline
\textbf{Key/Signature Reuse} & Incorrect implementation may reuse nonces or keys, weakening PQC scheme guarantees. & Use verified libraries (e.g., liboqs), follow NIST-compliant API usage, and enforce nonce uniqueness~\cite{dukhovni2023nonce}. & Medium \\
\hline
\textbf{Interoperability Failures} & Mixed deployments may fail due to unsupported PQC or hybrid modes in client or middleboxes. & Conduct fallback testing, negotiate cipher suite compatibility, and monitor DNS path transparency~\cite{ietf2023interop}. & Medium \\
\hline
\end{tabular}
\end{table}


\section{Experimental Evaluations}\label{sec:results}

In this section, we describe our experimental setup, metrics, numerical findings, and provide our analysis on the results. 

\subsection{Experimental Setup}

\begin{figure}
    \centering
    \includegraphics[width=0.5\linewidth]{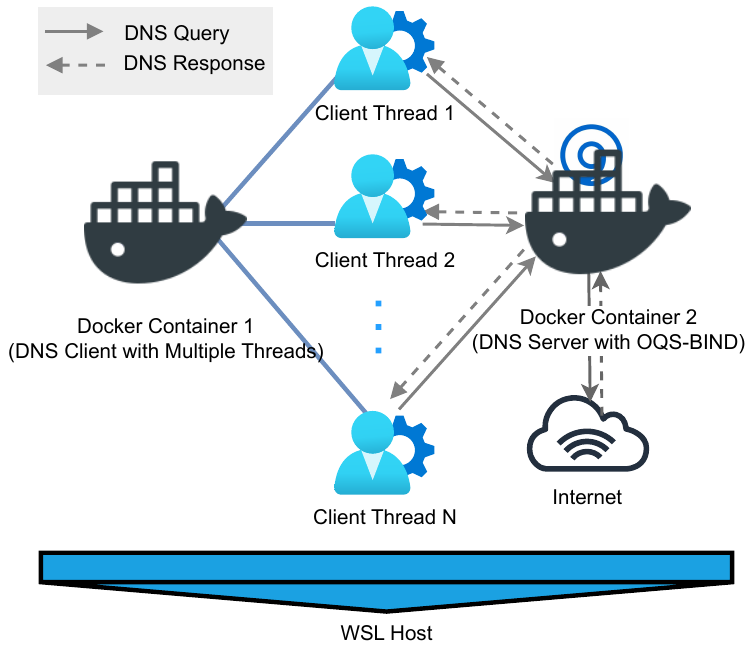}
    \caption{
    Experimental setup used in our evaluations.}
    \label{fig:experiment}
\end{figure}

The experimental testbed consisted of two Docker containers running on Ubuntu 22.04, deployed within a Windows Subsystem for Linux 2 (WSL2) as presented in Fig. \ref{fig:experiment} to isolate resource usage from host system activities and obtain accurate performance benchmarks. One container operated as a PQC-enabled local DNS resolver, while the other functioned as a client issuing DNS queries. The resolver was built using a forked version of BIND9 (OQS-BIND) ~\cite{OQS-Bind}, compiled with OpenSSL integrated with the Open Quantum Safe (OQS) library and \texttt{oqsprovider}. This configuration enabled support for post-quantum key exchange and digital signature algorithms. The resolver acted as a local DNS server capable of handling DNS-over-TLS (DoT), DNS-over-HTTPS (DoH), and DNSSEC queries. The client container generated DNS queries over both DoT and DoH protocols using the \texttt{dig} tool with the \texttt{+tls} and \texttt{+https} options, respectively. Each test run issued 100 queries to a test domain hosted on the resolver. 

The benchmark tests were conducted on a system running Windows 11 Home, equipped with an Intel(R) Core(TM) i7-10700 CPU operating at 2.90 GHz. The processor consists of 8 physical cores and 16 threads. The machine has 16.0 GB of installed RAM, of which 15.7 GB is usable. The experimental environment was set up using Windows Subsystem for Linux 2 (WSL2), with Ubuntu 22.04 running as the guest operating system. WSL2 dynamically allocated approximately 7.6 GiB of RAM for the guest system, and provided access to all 16 logical CPUs (8 cores with 2 threads per core). The software stack consisted of Docker version 26.1.3 as the container engine, OpenSSL 3.4.0 with integrated support for post-quantum cryptography, and liboqs version 0.12.0. The OQS OpenSSL provider used was oqsprovider version 0.8.0. The DNS resolver used in the experiments was a customized build of BIND, based on a forked version of BIND 9.19.17 (referred to as OQS-BIND), compiled to support PQC-DNS with liboqs and oqsprovider.



\subsection{Evaluation Framework and Benchmarking Metrics}
\label{sec:definition-benchmarking}

\begin{figure*}[htp!]
    \centering
    \includegraphics[width=1\linewidth]{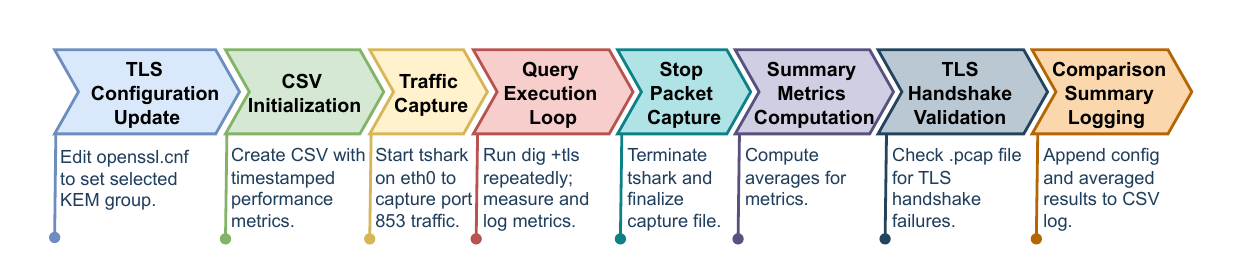}
    \caption{Process Steps of Benchmarking}
    \label{fig:processbar}
\end{figure*}

To evaluate the impact of different cryptographic configurations on DNS performance, we analyze the components of the performance vector \(\mathcal{S}_i(k, s)\), previously introduced in Section~\ref{ss:prob}. This vector represents the empirical behavior of the system under each KEM-signature pair \((k, s)\). 
We develop Python-based benchmarking scripts that automates DNS query execution, monitors system resources, captures network traffic, and aggregates performance data across repeated trials. 
It uses tools such as \texttt{psutil}, \texttt{/usr/bin/time}, and \texttt{tshark} to gather reproducible system-level measurements. The full benchmarking suite and some sample packet captures are available on GitHub\footnote{\url{https://github.com/ljy4499/pqc-dns}}. The step-by-step execution of the evaluation framework is formally defined in Algorithm \ref{alg1} and a high-level summary is presented in Fig. \ref{fig:processbar}. 
The individual components of \(\mathcal{S}_i(k, s)\) are defined as follows:


\begin{algorithm}[htp!]
\caption{Post-Quantum Secure DNS Evaluation}
\label{alg1}
\begin{algorithmic}[1]

\Procedure{EvaluateSecureDNS}{$\mathsf{KEM}, \mathsf{DS}, \mathsf{Domain}, N$}

  \State \textbf{// Step 1: TLS Configuration}
  \State Open $\mathsf{openssl.cnf}$
  \State Locate section $\mathsf{[system\_default\_sect]}$
  \State Update $\mathsf{Groups}$ parameter to $\mathsf{KEM}$

  \State \textbf{// Step 2: Metrics CSV Initialization}
  \State Create CSV with columns:
  \Statex \hspace{1.2em} \{$t$, $l$, $b$, $c$, $r$\}, where:
  \Statex \hspace{1.2em} $t$ = timestamp,\quad $l$ = latency (ms)
  \Statex \hspace{1.2em} $b$ = bandwidth (KB),\quad $c$ = CPU (\%),\quad $r$ = RAM (\%)

  \State \textbf{// Step 3: Start Traffic Capture}
  \State Launch $\mathsf{tshark}$ on interface $\mathsf{eth0}$
  \State Apply filter: $\mathsf{tcp\ port\ 853}$
  \State Save output to file $\mathsf{dot\_\{KEM\}\_\{DS\}.pcapng}$

  \State \textbf{// Step 4: Execute DNS Queries}
  \For{$i \gets 1$ to $N$}
    \State $B_{\text{pre}} \gets$ Network I/O snapshot
    \State $T_{\text{start}} \gets$ System time
    \State Execute $\mathsf{dig\ +tls\ @DS\ Domain}$
    \State $T_{\text{end}} \gets$ System time
    \State $B_{\text{post}} \gets$ Network I/O snapshot

    \State $l \gets (T_{\text{end}} - T_{\text{start}}) \times 1000$
    \State $b \gets (B_{\text{post}} - B_{\text{pre}}) / 1024$
    \State $c \gets$ CPU usage from \texttt{/usr/bin/time}
    \State $r \gets$ Peak RSS divided by container memory

    \State Append $\{t, l, b, c, r\}$ to CSV
  \EndFor

  \State \textbf{// Step 5: Stop Traffic Capture}
  \State Terminate $\mathsf{tshark}$ process

  \State \textbf{// Step 6: Summary Metric Computation}
  \State Read all records from CSV
  \State Compute: $\bar{l}$, $\bar{b}$, $\bar{c}$, $\bar{r}$
  \State Display summary statistics

  \State \textbf{// Step 7: TLS Handshake Validation}
  \State Run $\mathsf{tshark -r\ <pcap\_file>}$
  \If{``failure'' in output}
    \State Report TLS handshake error
  \Else
    \State Report handshake success
  \EndIf

  \State \textbf{// Step 8: Log Final Results}
  \State Append $\{\mathsf{KEM}, \mathsf{DS}, \bar{l}, \bar{b}, \bar{c}, \bar{r}\}$ to:
  \Statex \hspace{1.2em} $\mathsf{dot\_comparison\_results.csv}$

\EndProcedure

\end{algorithmic}
\end{algorithm}

\begin{itemize}[leftmargin=*]
  \item \textit{Latency} \((T_{\text{latency}})\): This value is calculated based on the time (in milliseconds) it takes between the first packet for a DNS request from the client to the last packet received for that request. This metric measures the responsiveness of the protocol and the ease of use.

  \item \textit{Bandwidth Usage} \((B_{\text{bandwidth}})\): The total network traffic generated by each DNS query, measured in kilobytes, taking into account both transmitted and received data. This reflects protocol efficiency and cryptographic overhead.

  \item \textit{CPU Usage} : The percentage of CPU time consumed during the DNS query was measured independently for the client and the server to assess the cryptographic workload on both sides:
 
  \textit{(i) Client CPU Usage} \((C_{\text{client}})\): This metric reflects the CPU utilisation of the \texttt{dig} process that initiates the DNS query. The \texttt{/usr/bin/time -v} utility was used to run \texttt{dig} and provides detailed statistics. The field ``Percent of CPU this job got'' was extracted using regular expressions. This raw percentage represents the ratio of CPU time to real (wall clock) time and is calculated internally as follows:
      \[
      \text{Raw CPU \%} = \left( \frac{\text{User Time} + \text{System Time}}{\text{Elapsed Time}} \right) \times 100\
      \]
  ``User Time'' accounts for execution of user-level instructions (e.g., cryptographic operations), while ``System Time'' includes time spent in kernel-level operations such as I/O. "Elapsed Time" refers to the total real-world (wall-clock) duration from the start to the end of the operation. To present this on a normalized 0–100\% scale, the raw value was divided by the number of virtual CPUs allocated to the container (\textbf{16 vCPUs}):
      \[
      \text{Normalized CPU \%} = \frac{\text{Raw CPU \%}}{n_{\text{vCPU}}}
      \]
   This normalization ensures that a value of 100\% represents full utilization of all assigned cores.
   
    \textit{(ii) Server CPU Usage} \(( C_{\text{server}})\): This metric was derived from periodic snapshots using \texttt{docker stats}, which provides live CPU utilization for containers. All other background processes were terminated to ensure the accuracy of the measurement.
   
       A monitoring script continuously sampled the container's CPU usage, and the highest observed percentage over the entire duration of client requests was recorded as the server's CPU usage.
       Internally, \texttt{docker stats} computes CPU usage as:
      \[
      \text{Raw CPU} = \left( \frac{\Delta \text{Container CPU \%}}{\Delta \text{System CPU \%}} \right) \times \text{vCPUs} \times 100
      \]
      \begin{itemize}
        \item $\Delta \text{Container CPU usage}$: The total amount of CPU time consumed by the container during the measurement period.
        \item $\Delta \text{System CPU usage}$: The total CPU time in the system for the same period.
        \item \texttt{vCPUs}: The number of virtual CPUs allocated to the container. In this case, the container was allocated \textbf{16 vCPUs}.
      \end{itemize}
       This formula calculates the proportion of CPU time consumed by the container relative to the host system's total CPU time, scaled by the number of virtual CPUs allocated to the container.
       To present this on a normalized 0–100\% scale, the raw value was divided by the number of virtual CPUs allocated to the container (\textbf{16 vCPUs}):
      \[
      \text{Normalized CPU \%} = \frac{\text{Raw CPU \%}}{n_{\text{vCPU}}}
      \]
       This adjustment ensures comparability with the normalized client CPU usage and expresses container CPU consumption as a proportion of its full computational capacity.

  \item \textit{Memory Usage} \((M_{\text{client}})\): The peak physical memory used by the \texttt{dig} process, relative to the total available memory in the container.
     The maximum resident set size (RSS) in kilobytes was obtained from \texttt{/usr/bin/time -v}.
     It was converted to mebibytes (MiB):
    \[
    \text{RAM Peak (MiB)} = \frac{\text{RSS (KB)}}{1024}
    \]
     RAM usage was computed as a percentage of total memory (i.e., 7810.3 MiB):
    \[
    \text{RAM Usage (\%)} = \left( \frac{\text{RAM Peak (MiB)}}{7810.3} \right) \times 100
    \]

\end{itemize}

\begin{algorithm} [htp!]
\caption{Execute Multi-client DNS Test Session}
\label{algmulti}
\begin{algorithmic}[1]
\Function{RunSession}{\textit{session\_id}}
    \State $B_{\text{start}} \gets$ \Call{GetTotalNetworkBytes}{}
    \State $T_{\text{start}} \gets$ \Call{GetCurrentTime}{}
    
    \State Initialize thread pool with 100 workers
    \ForAll{worker $\in$ thread pool}
        \State Submit \Call{RunDNSQuery}{}
    \EndFor
    
    \State Wait for all queries to complete
    
    \State $T_{\text{end}} \gets$ \Call{GetCurrentTime}{}
    \State $B_{\text{end}} \gets$ \Call{GetTotalNetworkBytes}{}
    
    \State $\Delta T \gets (T_{\text{end}} - T_{\text{start}}) \times 1000$ \Comment{Latency in ms}
    \State $\Delta B \gets (B_{\text{end}} - B_{\text{start}}) / 1024$ \Comment{Bandwidth in KB}
    
    \State \Call{WriteToCSV}{$session\_id$, $\Delta T$, $\Delta B$}
\EndFunction
\end{algorithmic}
\end{algorithm}


To evaluate the \textbf{scalability} and impact of PQC on resources with concurrent DNS workloads, a multi-threaded test configuration was used. In contrast to the single-threaded tests described above, this benchmark included 100 parallel workers (i.e., DNS clients), each sending one query per session, resulting in a total number of 10{,}000 queries in 100 sessions.  The Docker built-in utility \texttt{docker stats} was used to capture CPU and memory usage metrics. Resource utilisation was recorded by taking snapshots at regular intervals and extracting the peak values from the client and server during the test period. All reported CPU usage values were normalised on a scale from 0 to 100\%, which corresponds to the normalisation approach used in the single-thread evaluation. This ensures fair comparisons between different cryptographic methods and system configurations.
This benchmark specifically aims to evaluate the behaviour of PQC algorithms under load compared to traditional cryptographic methods, with a focus on throughput scaling and system resource evolution under high query volume.  Network usage and latency are measured as shown in Algorithm \ref{algmulti}.

\subsection{Experimental Results: Legacy vs. PQC Algorithms}

This section presents the benchmark results of DNS using various combinations of classical and post-quantum \ac{KEM} and \glspl{DS} at different NIST security levels \cite{nist:sp800-152}. For each evaluated cryptographic configuration \((k, s)\), we compute the corresponding performance vector \(\mathcal{S}_i(k, s)\) as defined in Section \ref{ss:prob}. The performance data are organized by protocol category, with separate tables for DoT, DoH, DNSSEC-enabled configurations ans security levels. All tables present results across the three cryptographic profiles: (i) Legacy-only, (ii) PQC-only and (iii) Hybrid.  
Each table includes measurements for latency, bandwidth usage, CPU utilization by the client/server, and memory consumption, whose notations and definitions are introduced earlier in the notation summary (Table~\ref{tab:notation}) and Section \ref{sec:definition-benchmarking}. We \textbf{bolded} the lowest numerical values in each subcategory for each metric in each table provided. Each combination \((k, s)\) is categorized based on the cryptographic primitives used to facilitate comparison across hybrid and fully post-quantum configurations. 

\begin{table}[htp!]
\footnotesize
  \caption{DNS over TLS Benchmark Results by Algorithms (Security Level 1)} 
  \vspace{-0.5em}
  \label{tab:sl1}
  \begin{tabular}{llrrrr}
  \toprule
  \textbf{KEM} & \textbf{DS} & \textbf{Latency(ms)} & \textbf{Bandwidth(kB)} & \textbf{Client / Server CPU(\%)} & 
  \textbf{Memory(\%)} \\
  \midrule
  \multicolumn{6}{l}{\underline{\textit{[Legacy(KEM) + Legacy(DS) Algorithms]}}} \\[0.25em]
  ffdhe2048 & rsa2048 &  10.08 & 4.14 & \textbf{5.00} / 0.95 & 0.151 \\
  ffdhe2048 & ecdsa-p256 & 9.67 & 3.58 & 5.39 / \textbf{0.71} & 0.154 \\
  ffdhe2048 & ed25519 & 9.61 & \textbf{3.50} & 5.38 / \textbf{0.71} & 0.153 \\
  secp256r1 & rsa2048 & 9.36 & 3.78 & 5.33 / 0.78 & 0.155 \\
  x25519 & rsa2048 & \textbf{9.17} & 3.71 & 5.27 / 0.76 & 0.153 \\[0.5em]
  \multicolumn{6}{l}{\underline{\textit{[PQC(KEM) + PQC(DS) Algorithms]}}} \\[0.25em]
  mlkem512 & mldsa44 & \textbf{9.10} & 10.54 & 5.80 / \textbf{0.50} & 0.165 \\
  mlkem512 & falcon512 & 9.11 & \textbf{6.54} & 5.47 / 0.57 & 0.164 \\
  mlkem512 & sphincssha2128f & 16.36 & 38.24 & \textbf{3.14} / 2.76 & 0.165 \\
  hqc128 & mldsa44 & 19.07 & 15.54 & 4.41 / 1.62 & 0.166 \\
  hqc128 & falcon512 & 19.13 & 11.67 & 4.39 / 1.64 & 0.164 \\[0.5em]
  \multicolumn{6}{l}{\underline{\textit{[Legacy(KEM) + PQC(DS) Algorithms]}}} \\[0.25em]
  ffdhe2048 & mldsa44 & 9.61 & 9.51 & 5.37 / 0.75 & 0.161 \\
  ffdhe2048 & falcon512 & 9.73 & \textbf{5.51} & 5.35 / 0.81 & 0.159 \\
  ffdhe2048 & sphincssha2128f & 17.16 & 37.19 & \textbf{3.15} / 2.81 & 0.158 \\
  secp256r1 & mldsa44 & \textbf{8.91} & 9.14 & 5.71 / 0.55 & 0.163 \\
  x25519 & falcon512 & 8.97 & 5.08 & 5.61 / \textbf{0.61} & 0.158 \\[0.5em]
  \multicolumn{6}{l}{\underline{\textit{[PQC(KEM) + Legacy(DS) Algorithms]}}} \\[0.25em]
  mlkem512 & rsa2048 & 9.37 & 5.17 & 5.42 / 0.74 & 0.161 \\
  mlkem512 & ecdsa-p256 & 8.96 & 4.60 & 5.77 / \textbf{0.47} & 0.163 \\
  mlkem512 & ed25519 & \textbf{8.91} & \textbf{4.54} & 5.90 / \textbf{0.47} & 0.161 \\
  hqc128 & rsa2048 & 19.67 & 10.17 & \textbf{4.37} / 1.71 & 0.160 \\
  hqc128 & ecdsa-p256 & 19.08 & 9.60 & 4.46 / 1.61 & 0.162 \\[0.5em]
  \bottomrule
  \end{tabular}
\end{table}

\begin{table}[htp!]
\footnotesize
  \caption{DNS over HTTPS Benchmark Results by Algorithms (Security Level 1)} 
  \vspace{-0.5em}
  \label{tab:sl1_doh}
  \begin{tabular}{llrrrr}
  \toprule
  \textbf{KEM} & \textbf{DS} & \textbf{Latency(ms)} & \textbf{Bandwidth(kB)} & \textbf{Client / Server CPU(\%)} & \textbf{Memory(\%)} \\
  \midrule
  \multicolumn{6}{l}{\underline{\textit{[Legacy(KEM) + Legacy(DS) Algorithms]}}} \\[0.25em]
  ffdhe2048 & rsa2048 & 9.97 & 4.54 & 5.00 / 0.97 & 0.152 \\
  ffdhe2048 & ecdsa-p256 & 9.67 & 3.96 & 5.41 / \textbf{0.74} & 0.155 \\
  ffdhe2048 & ed25519 & 9.63 & \textbf{3.90} & 5.40 / \textbf{0.74} & 0.154 \\
  secp256r1 & rsa2048 & 9.43 & 4.13 & \textbf{5.37} / 0.80 & 0.156 \\
  x25519 & rsa2048 & \textbf{9.19} & 4.13 & 5.39 / 0.80 & 0.154 \\[0.5em]
  \multicolumn{6}{l}{\underline{\textit{[PQC(KEM) + PQC(DS) Algorithms]}}} \\[0.25em]
  mlkem512 & mldsa44 & \textbf{8.93} & 10.95 & 5.76 / 0.52 & 0.166 \\
  mlkem512 & falcon512 & 9.07 & \textbf{6.89} & 5.53 / \textbf{0.60} & 0.166 \\
  mlkem512 & sphincssha2128f & 16.44 & 38.67 & \textbf{3.19} / 2.83 & 0.166 \\
  hqc128 & mldsa44 & 19.23 & 15.96 & 4.46 / 1.66 & 0.167 \\
  hqc128 & falcon512 & 19.11 & 12.10 & 4.41 / 1.67 & 0.165 \\[0.5em]
  \multicolumn{6}{l}{\underline{\textit{[Legacy(KEM) + PQC(DS) Algorithms]}}} \\[0.25em]
  ffdhe2048 & mldsa44 & 9.96 & 9.88 & 5.35 / 0.77 & 0.162 \\
  ffdhe2048 & falcon512 & 9.93 & 5.89 & 5.22 / 0.83 & 0.160 \\
  ffdhe2048 & sphincssha2128f & 17.11 & 37.64 & \textbf{3.16} / 2.91 & 0.159 \\
  secp256r1 & mldsa44 & \textbf{8.93} & 9.48 & 5.72 / \textbf{0.57} & 0.164 \\
  x25519 & falcon512 & 9.16 & \textbf{5.47} & 5.58 / 0.66 & 0.159 \\[0.5em]
  \multicolumn{6}{l}{\underline{\textit{[PQC(KEM) + Legacy(DS) Algorithms]}}} \\[0.25em]
  mlkem512 & rsa2048 & 9.41 & 5.60 & 5.36 / 0.77 & 0.162 \\
  mlkem512 & ecdsa-p256 & 8.87 & 5.00 & 5.84 / \textbf{0.51} & 0.163 \\
  mlkem512 & ed25519 & \textbf{8.84} & \textbf{4.88} & 5.93 / 0.49 & 0.161 \\
  hqc128 & rsa2048 & 19.49 & 10.57 & \textbf{4.41} / 1.74 & 0.161 \\
  hqc128 & ecdsa-p256 & 19.06 & 10.04 & 4.48 / 1.64 & 0.163 \\[0.5em]
  \bottomrule
  \end{tabular}
\end{table}

In Table \ref{tab:sl1}, the benchmark evaluates \ac{DoT} performance using a range of \ac{PQC} and legacy algorithms at NIST security level 1, which corresponds to the legacy 128-bit security level. 
Despite initial expectations that PQC algorithms would incur significant performance penalties relative to legacy counterparts, certain combinations-particularly \texttt{MLKEM512} paired with either \texttt{MLDSA44} or \texttt{Falcon512}-demonstrated latency metrics that closely matched or even outperformed legacy configurations. These combinations achieved latencies around 9 ms, which is comparable to commonly used Elliptic Curve and RSA-based solutions. However, a consistent trend was observed in bandwidth usage: PQC-based configurations generally incurred a higher data overhead. This is attributable to the larger key and signature sizes characteristic of quantum-resistant primitives. Among PQC candidates, \texttt{SPHINCS+-SHA2-128f} and \texttt{HQC-128} exhibited significantly higher latency and bandwidth usage. These results highlight a trade-off between post-quantum robustness and operational efficiency, making these algorithms less practical for latency-sensitive applications.

The benchmark in Table \ref{tab:sl1_doh} represents the performance of \ac{DoH} at NIST security level 1 and is directly comparable to the previously presented DoT benchmark results. For all analysed combinations of key exchange and digital signature, the metrics for latency, CPU usage and memory consumption show negligible differences between DoH and DoT. The main deviation lies in the bandwidth utilisation: DoH configurations consistently caused a slight overhead of 0.3 to 0.4 kilobytes on average compared to their DoT counterparts. This increase is due to the HTTP-based encapsulation and the additional protocol headers inherent to DoH.


The results in the Table \ref{tab:sl3} show the DoT performance results at NIST security level 3, which corresponds to 192 bits of security. One notable observation is the superior performance of \texttt{MLKEM768} paired with \texttt{MLDSA65}, which outperforms all tested legacy combinations in terms of latency, achieving the lowest measured value (9.24~ms). Compared to traditional key exchange mechanisms like \texttt{FFDHE3072} or \texttt{SECP384R1} paired with \texttt{RSA3072} or \texttt{ECDSA-P384}, this PQC combination delivers faster DoT performance. However, PQC algorithms exhibit a significantly higher bandwidth cost, consuming between \textbf{2$\times$ to 23$\times$} more bandwidth than legacy counterparts. For example, the \texttt{HQC192} with \texttt{SPHINCS+} signatures combination notably increases total bandwidth usage, with values exceeding 87~kB compared to sub-5~kB values in legacy configurations. This illustrates the tradeoff between post-quantum security and communication overhead, which must be considered when evaluating PQC deployment in bandwidth-sensitive environments.

\begin{table}[htp!]
\footnotesize
  \caption{DNS over TLS Benchmark Results by Algorithms (Security Level 3)} 
   \label{tab:sl3}
  \vspace{-0.5em}
  \begin{tabular}{llrrrr}
  \toprule
  \textbf{KEM} & \textbf{DS} & \textbf{Latency (ms)} & \textbf{Bandwidth (kB)} & \textbf{Client / Server CPU (\%)} & \textbf{Memory (\%)} \\
  \midrule
  \multicolumn{6}{l}{\underline{\textit{[Legacy(KEM) + Legacy(DS) Algorithms]}}} \\[0.25em]
  ffdhe3072 & rsa3072 & 12.51 & 4.77 & 4.45 / 1.58 & 0.152 \\
  ffdhe3072 & ecdsa-p384 & \textbf{12.29} & 3.92 & 4.62 / \textbf{1.25} & 0.152 \\
  secp384r1 & rsa3072 & 13.20 & 4.21 & \textbf{4.28} / 1.66 & 0.152 \\
  secp384r1 & ecdsa-p384 & 13.01 & \textbf{3.36} & 4.73 / 1.33 & 0.153 \\[0.5em]
  \multicolumn{6}{l}{\underline{\textit{[PQC(KEM) + PQC(DS) Algorithms]}}} \\[0.25em]
  mlkem768 & mldsa65 & \textbf{9.24} & \textbf{13.58} & 5.63 / \textbf{0.55} & 0.161 \\
  mlkem768 & sphincssha2192f & 21.69 & 75.63 & \textbf{2.50} / 3.49 & 0.162 \\
  hqc192 & mldsa65 & 39.64 & 24.68 & 3.95 / 2.19 & 0.166 \\
  hqc192 & sphincssha2192f & 51.95 & 86.63 & 3.11 / 3.01 & 0.164 \\[0.5em]
  \multicolumn{6}{l}{\underline{\textit{[Legacy(KEM) + PQC(DS) Algorithms]}}} \\[0.25em]
  ffdhe3072 & mldsa65 & \textbf{11.22} & 12.12 & 5.02 / \textbf{1.06} & 0.161 \\
  ffdhe3072 & sphincssha2192f & 23.47 & 74.16 & 2.47 / 3.46 & 0.158 \\
  secp384r1 & mldsa65 & 11.79 & \textbf{11.57} & 4.92 / 1.15 & 0.162 \\
  secp384r1 & sphincssha2192f & 24.17 & 73.61 & \textbf{2.45} / 3.46 & 0.159 \\[0.5em]
  \multicolumn{6}{l}{\underline{\textit{[PQC(KEM) + Legacy(DS) Algorithms]}}} \\[0.25em]
  mlkem768 & rsa3072 & 10.81 & 6.23 & 4.74 / 1.24 & 0.160 \\
  mlkem768 & ecdsa-p384 & \textbf{10.19} & \textbf{5.38} & 4.96 / \textbf{0.85} & 0.161 \\
  hqc192 & rsa3072 & 41.43 & 17.33 & \textbf{3.82} / 2.28 & 0.160 \\
  hqc192 & ecdsa-p384 & 41.05 & 16.48 & 3.92 / 2.20 & 0.161 \\[0.5em]
  \bottomrule
  \end{tabular}
\end{table}

The benchmark in Table \ref{tab:sl3_doh} shows the DoH performance results at NIST security level 3. The results show no significant difference compared to the previous DoT benchmark results. 

\begin{table}[htp!]
\footnotesize
  \caption{DNS over HTTPS Benchmark Results by Algorithms (Security Level 3)} 
  \label{tab:sl3_doh}
  \vspace{-0.5em}
  \begin{tabular}{llrrrr}
  \toprule
  \textbf{KEM} & \textbf{DS} & \textbf{Latency (ms)} & \textbf{Bandwidth (kB)} & \textbf{Client / Server CPU (\%)} & \textbf{Memory (\%)} \\
  \midrule
  \multicolumn{6}{l}{\underline{\textit{[Legacy(KEM) + Legacy(DS) Algorithms]}}} \\[0.25em]
  ffdhe3072 & rsa3072 & 12.49 & 5.12 & 4.44 / 1.61 & 0.153 \\
  ffdhe3072 & ecdsa-p384 & \textbf{12.36} & 4.34 & 4.69 / \textbf{1.33} & 0.154 \\
  secp384r1 & rsa3072 & 13.11 & 4.56 & \textbf{4.28} / 1.67 & 0.154 \\
  secp384r1 & ecdsa-p384 & 12.95 & \textbf{3.75} & 4.77 / 1.37 & 0.153 \\[0.5em]
  \multicolumn{6}{l}{\underline{\textit{[PQC(KEM) + PQC(DS) Algorithms]}}} \\[0.25em]
  mlkem768 & mldsa65 & \textbf{9.02} & \textbf{13.95} & 5.66 / \textbf{0.57} & 0.162 \\
  mlkem768 & sphincssha2192f & 21.67 & 76.06 & \textbf{2.50} / 3.51 & 0.163 \\
  hqc192 & mldsa65 & 39.79 & 25.13 & 3.97 / 2.24 & 0.167 \\
  hqc192 & sphincssha2192f & 51.99 & 87.07 & 3.13 / 3.02 & 0.165 \\[0.5em]
  \multicolumn{6}{l}{\underline{\textit{[Legacy(KEM) + PQC(DS) Algorithms]}}} \\[0.25em]
  ffdhe3072 & mldsa65 & \textbf{11.08} & 12.47 & 5.12 / \textbf{1.06} & 0.162 \\
  ffdhe3072 & sphincssha2192f & 23.60 & 74.59 & \textbf{2.47} / 3.46 & 0.159 \\
  secp384r1 & mldsa65 & 11.82 & \textbf{11.91} & 4.89 / 1.19 & 0.163 \\
  secp384r1 & sphincssha2192f & 24.25 & 74.05 & 2.49 / 3.46 & 0.160 \\[0.5em]
  \multicolumn{6}{l}{\underline{\textit{[PQC(KEM) + Legacy(DS) Algorithms]}}} \\[0.25em]
  mlkem768 & rsa3072 & 10.53 & 6.67 & 4.81 / 1.32 & 0.161 \\
  mlkem768 & ecdsa-p384 & \textbf{10.37} & \textbf{5.78} & 5.10 / \textbf{0.87} & 0.162 \\
  hqc192 & rsa3072 & 41.09 & 17.77 & \textbf{3.82} / 2.32 & 0.161 \\
  hqc192 & ecdsa-p384 & 40.96 & 16.93 & 3.93 / 2.22 & 0.162 \\[0.5em]
  \bottomrule
  \end{tabular}
\end{table}

\begin{table}[htp!]
\footnotesize
  \caption{DNS over TLS Benchmark Results by Algorithms (Security Level 5)}
  \vspace{-0.5em}
  \label{tab:sl5}
  \begin{tabular}{llrrrr}
  \toprule
  \textbf{KEM} & \textbf{DS} & \textbf{Latency (ms)} & \textbf{Bandwidth (kB)} & \textbf{Client / Server CPU (\%)} & \textbf{Memory (\%)} \\
  \midrule
  \multicolumn{6}{l}{\underline{\textit{[Legacy(KEM) + Legacy(DS) Algorithms]}}} \\[0.25em]
  ffdhe4096 & rsa4096 & 16.65 & 5.39 & 3.73 / 2.26 & 0.152 \\
  ffdhe4096 & ecdsa-p521 & 16.56 & 4.26 & 4.17 / 1.76 & 0.153 \\
  ffdhe4096 & ed448 & 13.51 & \textbf{4.13} & 4.63 / \textbf{1.40} & 0.152 \\
  secp521r1 & rsa4096 & 18.74 & 4.66 & \textbf{3.59} / 2.37 & 0.153 \\
  x448 & rsa4096 & \textbf{12.47} & 4.51 & 3.89 / 2.02 & 0.152 \\[0.5em]
  \multicolumn{6}{l}{\underline{\textit{[PQC(KEM) + PQC(DS) Algorithms]}}} \\[0.25em]
  mlkem1024 & mldsa87 & \textbf{9.13} & 17.76 & 5.48 / \textbf{0.57} & 0.165 \\
  mlkem1024 & falcon1024 & 9.29 & \textbf{10.28} & 5.42 / 0.71 & 0.165 \\
  hqc256 & mldsa87 & 65.97 & 36.11 & 3.78 / 2.38 & 0.167 \\
  hqc256 & falcon1024 & 65.54 & 28.64 & \textbf{3.75} / 2.39 & 0.166 \\[0.5em]
  \multicolumn{6}{l}{\underline{\textit{[Legacy(KEM) + PQC(DS) Algorithms]}}} \\[0.25em]
  ffdhe4096 & mldsa87 & 13.54 & 15.70 & 4.63 / 1.41 & 0.162 \\
  ffdhe4096 & falcon1024 & 13.83 & 8.09 & 4.49 / 1.48 & 0.159 \\
  secp521r1 & mldsa87 & 16.64 & 14.95 & \textbf{4.28} / 1.65 & 0.162 \\
  secp521r1 & falcon1024 & 15.90 & 7.36 & 4.31 / 1.68 & 0.159 \\
  x448 & mldsa87 & \textbf{9.63} & 14.82 & 5.40 / \textbf{0.71} & 0.159 \\
  x448 & falcon1024 & 9.90 & \textbf{7.21} & 5.31 / 0.84 & 0.158 \\[0.5em]
  \multicolumn{6}{l}{\underline{\textit{[PQC(KEM) + Legacy(DS) Algorithms]}}} \\[0.5em]
  mlkem1024 & rsa4096 & 12.38 & 7.46 & 3.95 / 1.96 & 0.159 \\
  mlkem1024 & ecdsa-p521 & 13.07 & 6.30 & 4.41 / 1.24 & 0.158  \\
  mlkem1024 & ed448 & \textbf{9.13} & \textbf{6.19} & 5.44 / \textbf{0.56} & 0.158 \\
  hqc256 & rsa4096 & 68.78 & 25.93 & \textbf{3.60} / 2.54 & 0.161 \\
  hqc256 & ecdsa-p521 & 68.15 & 24.82 & 3.74 / 2.47 & 0.161 \\[0.5em]
  \bottomrule
  \end{tabular}
\end{table}

The benchmark in Table \ref{tab:sl5} shows the DoT performance results at NIST security level 5, which corresponds to a 256-bit security level of the legacy algorithms. One noticeable observation is the consistently low latency achieved by the PQC algorithms \texttt{MLKEM1024} with \texttt{MLDSA87}, achieving sub-10 ms latency. These results outperform several legacy combinations-such as \texttt{FFDHE4096} with \texttt{RSA4096}-which exhibited latencies ranging from 12 to 18 ms. However, not all PQC configurations demonstrated such efficiency. The \texttt{HQC256} key encapsulation mechanism, when paired with either PQC or legacy digital signature schemes, significantly increased the latency, reaching values over 65 ms. This degradation in performance is primarily due to HQC's large key and ciphertext sizes, which impose higher computational and transmission overhead during the TLS handshake process. 

\begin{figure}[htp!]
  \centering
  \begin{subfigure}[t]{0.49\linewidth}
    \includegraphics[width=\linewidth]{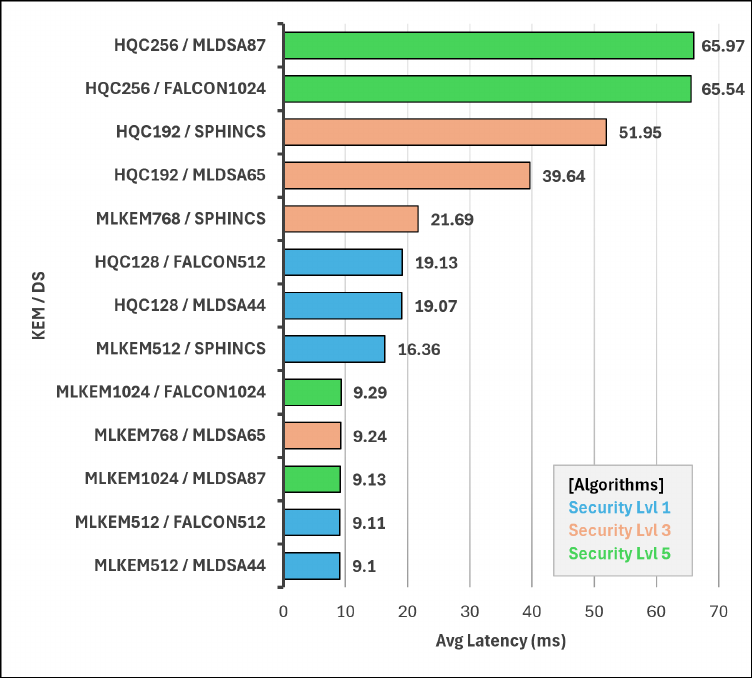}
    \Description{Bar chart showing DNS over TLS latency comparison by Security Level 1, 3, and 5} 
  \end{subfigure}
  \hfill
  \begin{subfigure}[t]{0.49\linewidth}
    \includegraphics[width=\linewidth]{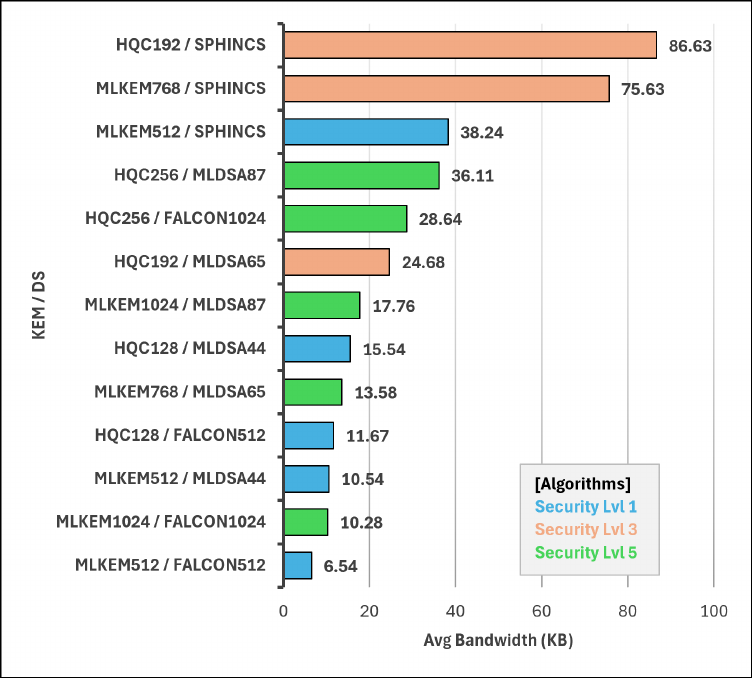}
    \Description{Bar chart showing DNS over TLS bandwidth comparison by Security Level 1, 3, and 5} 
  \end{subfigure}
  \vspace{-1em}
  \caption{DoT Latency (left) and Bandwidth(right) comparison chart by Security Level 1, 3, and 5}
  \label{fig:dot135}
\end{figure}

Fig. \ref{fig:dot135} shows a comparison of DoT performance at NIST security levels 1, 3 and 5. It is noteworthy that the \ac{KEM} of \texttt{MLKEM} in conjunction with the \ac{DS} algorithms of \texttt{MLDSA} or \texttt{Falcon} has a consistently low latency regardless of the security level, indicating good scalability of performance. In contrast, algorithms such as \texttt{HQC} and \texttt{SPHINCS+} showed a proportional increase in latency with increasing security level, reflecting their greater computational and structural overhead. In terms of bandwidth, all configurations showed a clear trend: as the security level increased, so did the bandwidth consumption. This is due to the larger key and ciphertext sizes required to maintain a higher level of cryptographic strength, especially for code-based or stateless hash-based schemes such as \texttt{HQC} and \texttt{SPHINCS+}.

The benchmark in Table \ref{tab:sl5_doh} presents DoH performance results at NIST security level 5. The results show no significant difference when compared to the previous DoT benchmark result. To avoid redundancy and enhance readability, a comparison chart is omitted. 

\begin{table}[htp!]
\footnotesize
  \caption{DNS over HTTPS Benchmark Results by Algorithms (Security Level 5)}
  \label{tab:sl5_doh}
  \vspace{-0.5em}
  \begin{tabular}{llrrrr}
  \toprule
  \textbf{KEM} & \textbf{DS} & \textbf{Latency (ms)} & \textbf{Bandwidth (kB)} & \textbf{Client / Server CPU (\%)} & \textbf{Memory (\%)} \\
  \midrule
  \multicolumn{6}{l}{\underline{\textit{[Legacy(KEM) + Legacy(DS) Algorithms]}}} \\[0.25em]
  ffdhe4096 & rsa4096 & 16.78 & 5.84 & 3.68 / 2.27 & 0.153 \\
  ffdhe4096 & ecdsa-p521 & 16.79 & 4.72 & 4.15 / 1.80 & 0.154 \\
  ffdhe4096 & ed448 & 13.63 & \textbf{4.60} & 4.64 / \textbf{1.43} & 0.153 \\
  secp521r1 & rsa4096 & 18.82 & 5.07 & \textbf{3.60} / 2.39 & 0.153 \\
  x448 & rsa4096 & \textbf{12.70} & 4.95 & 3.89 / 2.03 & 0.153 \\[0.5em]
  \multicolumn{6}{l}{\underline{\textit{[PQC(KEM) + PQC(DS) Algorithms]}}} \\[0.25em]
  mlkem1024 & mldsa87 & \textbf{9.41} & 18.17 & 5.43 / \textbf{0.59} & 0.166 \\
  mlkem1024 & falcon1024 & 9.50 & \textbf{10.67} & 5.38 / 0.74 & 0.166 \\
  hqc256 & mldsa87 & 65.52 & 36.58 & \textbf{3.77} / 2.41 & 0.168 \\
  hqc256 & falcon1024 & 65.89 & 29.10 & \textbf{3.77} / 2.46 & 0.167 \\[0.5em]
  \multicolumn{6}{l}{\underline{\textit{[Legacy(KEM) + PQC(DS) Algorithms]}}} \\[0.25em]
  ffdhe4096 & mldsa87 & 13.78 & 16.03 & 4.60 / 1.42 & 0.162 \\
  ffdhe4096 & falcon1024 & 13.95 & 8.50 & 4.46 / 1.54 & 0.160 \\
  secp521r1 & mldsa87 & 15.82 & 15.43 & 4.41 / 1.66 & 0.163 \\
  secp521r1 & falcon1024 & 16.00 & 7.76 & \textbf{4.27} / 1.73 & 0.160 \\
  x448 & mldsa87 & \textbf{9.75} & 15.17 & 5.39 / \textbf{0.73} & 0.162 \\
  x448 & falcon1024 & 10.04 & \textbf{7.62} & 5.19 / 0.86 & 0.159 \\[0.5em]
  \multicolumn{6}{l}{\underline{\textit{[PQC(KEM) + Legacy(DS) Algorithms]}}} \\[0.25em]
  mlkem1024 & rsa4096 & 12.36 & 7.86 & 3.90 / 1.97 & 0.160 \\
  mlkem1024 & ecdsa-p521 & 12.06 & 6.72 & 4.56 / 1.27 & 0.161 \\
  mlkem1024 & ed448 & \textbf{9.20} & \textbf{6.62} & 5.46 / \textbf{0.59} & 0.159 \\
  hqc256 & rsa4096 & 68.44 & 26.41 & \textbf{3.62} / 2.59 & 0.162 \\
  hqc256 & ecdsa-p521 & 68.17 & 25.28 & 3.74 / 2.46 & 0.162 \\[0.5em]
  \bottomrule
  \end{tabular}
\end{table}

This benchmark in Table \ref{tab:dnssec} presents the DNSSEC performance results. While PQC-based DNSSEC algorithms exhibit higher bandwidth usage compared to legacy algorithms, this increase is primarily due to the larger size of the cryptographic signatures and certificates. Despite this overhead, the latency remains nearly unaffected across all tested algorithms. For example, \texttt{MLDSA44}  and \texttt{SPHINCS+} variants show significantly higher bandwidth consumption than \texttt{RSA2048} or \texttt{ED25519}, yet maintain comparable latency values, indicating that DNSSEC verification remains computationally lightweight even with post-quantum algorithms. 

\begin{table}[htp!]
\footnotesize
  \caption{DNSSEC Benchmark Results of Algorithms}
  \label{tab:dnssec}
  \vspace{-0.5em}
  \begin{tabular}{lrrrr}
  \toprule
  \textbf{DNSSEC} & \textbf{Latency (ms)} & \textbf{Bandwidth (kB)} & \textbf{Client / Server CPU (\%)} & \textbf{Memory (\%)} \\
  \midrule
  \multicolumn{5}{l}{\underline{\textit{[Legacy(DNSSEC) Algorithms]}}} \\[0.25em]
  rsa2048sha256 & 6.22 & 0.53 & 6.08 / 0.10 & 0.141 \\
  ecdsap256sha256 & 6.28 & \textbf{0.34} & 6.11 / 0.10 & 0.141 \\
  ed25519 & \textbf{6.20} & \textbf{0.34} & \textbf{5.99} / 0.10 & 0.141 \\[0.5em]
  \multicolumn{5}{l}{\underline{\textit{[PQC(DNSSEC) Algorithms]}}} \\[0.25em]
  mldsa44 & 6.83 & 3.46 & 6.07 / 0.25 & 0.141 \\
  falconpadded512 & \textbf{6.29} & \textbf{0.93} & 6.16 / \textbf{0.10} & 0.141 \\
  sphincssha2128fsimple & 6.88 & 8.89 & \textbf{5.91} / 0.26 & 0.142 \\[0.5em]
  \bottomrule
  \end{tabular}
\end{table}

The benchmark in  Table \ref{tab:sl1_dnssec} presents the performance results for DNSSEC combined with DoT using a range of cryptographic algorithms at NIST Security Level 1. The overall behavior of the algorithms remains consistent with the non-DNSSEC configurations, particularly in terms of latency, CPU, and memory usage. The primary difference observed is an increase in bandwidth, which is expected due to the additional DNSSEC-related cryptographic material. However, this increase does not have a significant impact on latency or other system metrics. Algorithms such as \texttt{SPHINCS+}, which already produce large signature sizes, contribute substantially to the bandwidth overhead. Consequently, configurations involving \texttt{SPHINCS+} exhibit significantly higher bandwidth consumption compared to other algorithms. 

\begin{table}[ht]
\footnotesize
  \caption{DNSSEC with DoT Benchmark Results by Algorithms (Security Level 1)}
  \label{tab:sl1_dnssec}
  \vspace{-0.5em}
  \begin{tabular}{lllrrrr}
  \toprule
  \textbf{DNSSEC} & \textbf{KEM} & \textbf{DS} & \textbf{{\small Latency (ms)}} & \textbf{{\small Bandwidth (kB)}} & \textbf{{\small Client / Server CPU (\%)}} & \textbf{{\small Memory (\%)}} \\
  \midrule
  \multicolumn{7}{l}{\underline{\textit{[Legacy(DNSSEC) + Legacy(KEM) + Legacy(DS) Algorithms]}}} \\[0.25em]
  rsa2048 & ffdhe2048 & rsa2048 & 10.32 & 4.44 & \textbf{4.92} / 0.94 & 0.153 \\
  ecdsap256 & ffdhe2048 & ecdsa-p256 & 9.77 & 3.68 & 5.38 / 0.72 & 0.156 \\
  ed25519 & ffdhe2048 & ed25519 & \textbf{9.55} & \textbf{3.61} & 5.42 / 0.71 & 0.154 \\[0.5em]
  \multicolumn{7}{l}{\underline{\textit{[PQC(DNSSEC) + PQC(KEM) + PQC(DS) Algorithms]}}} \\[0.25em]
  mldsa44 & mlkem512 & mldsa44 & 9.21 & 12.95 & 5.55 / \textbf{0.51} & 0.166 \\
  falcon512 & mlkem512 & falcon512 & \textbf{9.05} & \textbf{7.24} & 5.51 / 0.59 & 0.165 \\
  {\scriptsize sphincssha2128f} & mlkem512 & {\scriptsize sphincssha2128f} & 16.31 & 45.96 & \textbf{3.18} / 2.79 & 0.165 \\[0.5em]
  \multicolumn{7}{l}{\underline{\textit{[Legacy(DNSSEC) + PQC(KEM) + PQC(DS) Algorithms]}}} \\[0.25em]
  rsa2048 & mlkem512 & mldsa44 & 9.04 & 10.84 & 5.83 / \textbf{0.50} & 0.166 \\
  ecdsap256 & mlkem512 & falcon512 & \textbf{9.02} & \textbf{6.66} & 5.54 / 0.58 & 0.165 \\
  ed25519 & mlkem512 & {\scriptsize sphincssha2128f} & 16.35 & 38.35 & \textbf{3.16} / 2.78 & 0.165 \\[0.5em]
  \multicolumn{7}{l}{\underline{\textit{[PQC(DNSSEC) + Legacy(KEM) + Legacy(DS) Algorithms]}}} \\[0.25em]
  mldsa44 & ffdhe2048 & rsa2048 & 10.24 & 6.55 & \textbf{4.95} / 0.94 & 0.153 \\
  falcon512 & ffdhe2048 & ecdsa-p256 & \textbf{9.69} & \textbf{4.27} & 5.34 / \textbf{0.71} & 0.156 \\
  {\scriptsize sphincssha2128f} & ffdhe2048 & ed25519 & 9.89 & 11.29 & 5.29 / \textbf{0.71} & 0.154 \\[0.5em]
  \bottomrule
  \end{tabular}
\end{table}

This benchmark in Table \ref{tab:sl1_dnssec_doh} presents DNSSEC with DoH performance results at NIST security level 1. The results show no significant difference when compared to the previous DoT benchmark result. To avoid redundancy and enhance readability, a comparison chart is omitted. 

\begin{table}[htp!]
\footnotesize
  \caption{DNSSEC with DoH Benchmark Results by Algorithms (Security Level 1)}
  \label{tab:sl1_dnssec_doh}
  \vspace{-0.5em}
  \begin{tabular}{lllrrrr}
  \toprule
  \textbf{DNSSEC} & \textbf{KEM} & \textbf{DS} & \textbf{{\small Latency (ms)}} & \textbf{{\small Bandwidth (kB)}} & \textbf{{\small Client / Server CPU (\%)}} & \textbf{{\small Memory (\%)}} \\
  \midrule
  \multicolumn{7}{l}{\underline{\textit{[Legacy(DNSSEC) + Legacy(KEM) + Legacy(DS) Algorithms]}}} \\[0.25em]
  rsa2048 & ffdhe2048 & rsa2048 & 10.88 & 4.86 & \textbf{4.93} / 0.99 & 0.154 \\
  ecdsap256 & ffdhe2048 & ecdsa-p256 & 9.72 & 4.07 & 5.40 / 0.76 & 0.157 \\
  ed25519 & ffdhe2048 & ed25519 & \textbf{9.56} & \textbf{4.02} & 5.42 / \textbf{0.74} & 0.155 \\[0.5em]
  \multicolumn{7}{l}{\underline{\textit{[PQC(DNSSEC) + PQC(KEM) + PQC(DS) Algorithms]}}} \\[0.25em]
 mldsa44 & mlkem512 & mldsa44 & 9.18 & 13.39 & 5.56 / \textbf{0.53} & 0.167 \\
  falconp512 & mlkem512 & falcon512 & \textbf{9.06} & \textbf{7.64} & 5.49 / 0.60 & 0.166 \\
  {\scriptsize sphincssha2128f} & mlkem512 & {\scriptsize sphincssha2128f} & 16.38 & 46.42 & \textbf{3.27} / 2.79 & 0.166 \\[0.5em]
  \multicolumn{7}{l}{\underline{\textit{[Legacy(DNSSEC) + PQC(KEM) + PQC(DS) Algorithms]}}} \\[0.25em]
  rsa2048 & mlkem512 & mldsa44 & \textbf{9.01} & 11.25 & 5.68 / \textbf{0.57} & 0.167 \\
  ecdsap256 & mlkem512 & falcon512 & 9.06 & \textbf{7.06} & 5.46 / 0.62 & 0.166 \\
  ed25519 & mlkem512 & {\scriptsize sphincssha2128f} & 16.46 & 38.78 & \textbf{3.25} / 2.78 & 0.166 \\[0.5em]
  \multicolumn{7}{l}{\underline{\textit{[PQC(DNSSEC) + Legacy(KEM) + Legacy(DS) Algorithms]}}} \\[0.25em]
  mldsa44 & ffdhe2048 & rsa2048 & 10.09 & 6.98 & \textbf{4.95} / 0.97 & 0.154 \\
  falcon512 & ffdhe2048 & ecdsa-p256 & \textbf{9.69} & \textbf{4.65} & 5.39 / 0.81 & 0.157 \\
  {\scriptsize sphincssha2128f} & ffdhe2048 & ed25519 & 9.76 & 11.74 & 5.39 / \textbf{0.74} & 0.155 \\[0.5em]
  \bottomrule
  \end{tabular}
\end{table}

The results in Table \ref{tab:sl1_100} list DoT performance under a multi-thread scenario using 100 concurrent queries at NIST Security Level 1. The results show consistent trends with earlier benchmarks: PQC key exchange mechanisms such as \texttt{MLKEM512} combined with signature schemes like \texttt{MLDSA44} or \texttt{Falcon512} yield comparable or even lower latencies than several legacy combinations. Notably, despite the significantly increased bandwidth-scaling linearly with the number of concurrent queries-the latency remains unaffected. This is likely due to the small size of each DNS query and response, which minimizes network congestion and avoids saturating the bandwidth capacity, especially under typical modern network conditions. With these observations, DNS performance confirms that latency is often dominated by cryptographic processing and handshake round-trips, not data volume per se. However, algorithms with heavier computational loads-such as \texttt{SPHINCS+} and \texttt{HQC}-introduce higher latencies, correlating with increased server-side CPU usage. The observed CPU impact suggests that for algorithms with large key or signature sizes, the processing overhead (e.g., signature verification or key decoding) becomes the dominant latency factor under load, rather than network throughput.

\begin{table}[htp!]
\footnotesize
  \caption{Multi-thread (100 Concurrent Queries) DNS over TLS Benchmark Results by Algorithms (Security Level 1)}
  \label{tab:sl1_100}
  \vspace{-0.5em}
  \begin{tabular}{llrrrr}
  \toprule
  \textbf{KEM} & \textbf{DS} & \textbf{Latency (ms)} & \textbf{Bandwidth (kB)} & \textbf{Client / Server CPU (\%)} & \textbf{Memory (\%)} \\
  \midrule
  \multicolumn{6}{l}{\underline{\textit{[Legacy(KEM) + Legacy(DS) Algorithms]}}} \\[0.25em]
  ffdhe2048 & rsa2048 & 125.28 & 412.28 & \textbf{72.01} / 13.18 & 2.66 \\
  ffdhe2048 & ecdsa-p256 & 111.55 & 354.97 & 74.71 / \textbf{9.61} & 2.09 \\
  ffdhe2048 & ed25519 & 113.53 & \textbf{347.88} & 75.07 / 9.67 & 3.03 \\
  secp256r1 & rsa2048 & 114.23 & 375.25 & 72.68 / 10.63 & 3.04 \\
  x25519 & rsa2048 & \textbf{109.89} & 368.80 & 73.32 / 10.44 & 2.93 \\[0.5em]
  \multicolumn{6}{l}{\underline{\textit{[PQC(KEM) + PQC(DS) Algorithms]}}} \\[0.25em]
  mlkem512 & mldsa44 & \textbf{106.08} & 1044.63 & 76.80 / \textbf{6.30} & 3.91 \\
  mlkem512 & falcon512 & 107.54 & \textbf{651.29} & 76.52 / 7.33 & 3.50 \\
  mlkem512 & sphincssha2128f & 206.10 & 3827.23 & \textbf{44.88} / 42.60 & 6.08 \\
  hqc128 & mldsa44 & 231.62 & 1553.43 & 65.56 / 26.24 & 3.74 \\
  hqc128 & falcon512 & 234.98 & 1167.13 & 64.88 / 26.46 & 3.59 \\[0.5em]
  \multicolumn{6}{l}{\underline{\textit{[Legacy(KEM) + PQC(DS) Algorithms]}}} \\[0.25em]
  ffdhe2048 & mldsa44 & 112.65 & 948.52 & 74.38 / 10.09 & 3.04 \\
  ffdhe2048 & falcon512 & 114.85 & \textbf{549.03} & 72.28 / 10.86 & 2.91 \\
  ffdhe2048 & sphincssha2128f & 215.98 & 3724.60 & \textbf{45.17} / 42.33 & 7.98 \\
  secp256r1 & mldsa44 & \textbf{105.98} & 907.35 & 76.50 / \textbf{7.13} & 3.94 \\
  x25519 & falcon512 & 109.67 & 505.50 & 76.00 / 7.70 & 3.45 \\[0.5em]
  \multicolumn{6}{l}{\underline{\textit{[PQC(KEM) + Legacy(DS) Algorithms]}}} \\[0.25em]
  mlkem512 & rsa2048 & 112.60 & 514.90 & 73.66 / 10.03 & 3.25 \\
  mlkem512 & ecdsa-p256 & \textbf{110.05} & 450.78 & 77.69 / 5.92 & 3.54 \\
  mlkem512 & ed25519 & 111.08 & \textbf{444.13} & 75.96 / \textbf{5.83} & 3.52 \\
  hqc128 & rsa2048 & 242.12 & 1017.43 & \textbf{63.62} / 27.73 & 3.78 \\
  hqc128 & ecdsa-p256 & 234.85 & 960.10 & 65.85 / 26.15 & 3.87 \\[0.5em]
  \bottomrule
  \end{tabular}
\end{table}

Meanwile, the results in Table \ref{tab:sl1_doh_100} presents Multi-thread (100 Concurrent Queries) DoH performance results at NIST security level 1. The results show no significant difference when compared to the previous DoT benchmark result. To avoid redundancy and enhance readability, a comparison chart is omitted. 

\begin{table}[htp!]
\footnotesize
  \caption{Multi-thread (100 Concurrent Queries) DNS over HTTPS Benchmark Results by Algorithms (Security Level 1)}
  \label{tab:sl1_doh_100}
  \vspace{-0.5em}
  \begin{tabular}{llrrrr}
  \toprule
  \textbf{KEM} & \textbf{DS} & \textbf{Latency (ms)} & \textbf{Bandwidth (kB)} & \textbf{Client / Server CPU (\%)} & \textbf{Memory (\%)} \\
  \midrule
  \multicolumn{6}{l}{\underline{\textit{[Legacy(KEM) + Legacy(DS) Algorithms]}}} \\[0.25em]
  ffdhe2048 & rsa2048 & 116.57 & 442.58 & \textbf{71.82} / 13.89 & 2.72 \\
  ffdhe2048 & ecdsa-p256 & 114.43 & 386.64 & 74.09 / 9.82 & 3.69 \\
  ffdhe2048 & ed25519 & 126.00 & \textbf{378.61} & 76.88 / \textbf{9.35} & 3.77 \\
  secp256r1 & rsa2048 & \textbf{113.69} & 406.72 & 73.16 / 11.06 & 3.16 \\
  x25519 & rsa2048 & 123.77 & 399.66 & 72.56 / 10.61 & 3.83 \\[0.5em]
  \multicolumn{6}{l}{\underline{\textit{[PQC(KEM) + PQC(DS) Algorithms]}}} \\[0.25em]
  mlkem512 & mldsa44 & \textbf{117.57} & 1085.28 & 78.26 / \textbf{6.85} & 3.18 \\
  mlkem512 & falcon512 & 128.04 & \textbf{683.36} & 79.03 / 7.34 & 3.63 \\
  mlkem512 & sphincssha2128f & 213.58 & 3857.46 & \textbf{44.32} / 42.27 & 6.47 \\
  hqc128 & mldsa44 & 231.99 & 1581.33 & 65.17 / 26.22 & 3.77 \\
  hqc128 & falcon512 & 236.47 & 1194.90 & 65.46 / 26.67 & 4.36 \\[0.5em]
  \multicolumn{6}{l}{\underline{\textit{[Legacy(KEM) + PQC(DS) Algorithms]}}} \\[0.25em]
  ffdhe2048 & mldsa44 & 126.36 & 979.29 & 73.72 / 10.34 & 4.17 \\
  ffdhe2048 & falcon512 & 127.36 & 579.55 & 72.97 / 11.28 & 3.97 \\
  ffdhe2048 & sphincssha2128f & 230.49 & 3754.98 & \textbf{46.35} / 42.37 & 5.07 \\
  secp256r1 & mldsa44 & \textbf{114.97} & 943.96 & 76.35 / \textbf{7.50} & 4.38 \\
  x25519 & falcon512 & 115.84 & \textbf{537.87} & 76.85 / 7.93 & 3.64 \\[0.5em]
  \multicolumn{6}{l}{\underline{\textit{[PQC(KEM) + Legacy(DS) Algorithms]}}} \\[0.25em]
  mlkem512 & rsa2048 & 125.30 & 545.89 & 75.47 / 10.22 & 3.68 \\
  mlkem512 & ecdsa-p256 & \textbf{116.64} & 491.30 & 78.84 / 6.33 & 3.50 \\
  mlkem512 & ed25519 & 129.77 & \textbf{483.09} & 79.06 / \textbf{6.10} & 4.02 \\
  hqc128 & rsa2048 & 240.28 & 1045.36 & \textbf{63.94} / 27.50 & 4.70 \\
  hqc128 & ecdsa-p256 & 225.07 & 987.77 & 65.90 / 26.06 & 3.98 \\[0.5em]
  \bottomrule
  \end{tabular}
\end{table}

Compared to the 100-concurrent-query benchmark, the 1000-query scenario presented in Table \ref{tab:sl1_1000} reveals important scalability distinctions between algorithmic combinations. While general latency trends remain consistent, the increased concurrency amplifies differences in resource handling, particularly CPU usage and bandwidth overhead. For example, combinations like \texttt{MLKEM512 + Falcon512} continue to perform efficiently, but now exhibit slightly higher client CPU utilization, reflecting the cumulative impact of repeated cryptographic operations at scale. Conversely, resource-intensive algorithms such as \texttt{HQC128 + SPHINCS+} and \texttt{HQC128 + Falcon512} demonstrate more pronounced performance degradation: latencies exceed 2200 ms and server CPU usage approaches 30 percent, signaling stress under high load. Bandwidth scales predictably with query volume, but bandwidth-intensive schemes-especially those using \texttt{SPHINCS+}-now produce transfer volumes exceeding 38 MB per session, reinforcing concerns about deployment in constrained networks. Notably, hybrid combinations involving PQC signatures (e.g., \texttt{ffdhe2048 + SPHINCS+}) show similar trends, with CPU bottlenecks becoming more evident than in the lower-concurrency test. 

\begin{table}[htp!]
\footnotesize
  \caption{Multi-thread (1000 Concurrent Queries) DNS over TLS Benchmark Results by Algorithms (Security Level 1)}
  \label{tab:sl1_1000}
  \vspace{-0.5em}
  \begin{tabular}{llrrrr}
  \toprule
  \textbf{KEM} & \textbf{DS} & \textbf{Latency (ms)} & \textbf{Bandwidth (kB)} & \textbf{Client / Server CPU (\%)} & \textbf{Memory (\%)} \\
  \midrule
  \multicolumn{6}{l}{\underline{\textit{[Legacy(KEM) + Legacy(DS) Algorithms]}}} \\[0.25em]
  ffdhe2048 & rsa2048 & 1170.67 & 4135.29 & \textbf{81.76} / 16.36 & 3.17 \\
  ffdhe2048 & ecdsa-p256 & 1128.47 & 3564.65 & 86.91 / 11.66 & 3.19 \\
  ffdhe2048 & ed25519 & 1138.10 & \textbf{3495.04} & 86.47 / \textbf{11.42} & 3.17 \\
  secp256r1 & rsa2048 & 1134.85 & 3766.40 & 85.14 / 12.80 & 3.03 \\
  x25519 & rsa2048 & \textbf{1038.85} & 3700.32 & 84.26 / 12.16 & 2.93 \\[0.5em]
  \multicolumn{6}{l}{\underline{\textit{[PQC(KEM) + PQC(DS) Algorithms]}}} \\[0.25em]
  mlkem512 & mldsa44 & 1041.92 & 10525.63 & 88.90 / \textbf{7.23} & 3.14 \\
  mlkem512 & falcon512 & \textbf{1015.81} & \textbf{6531.43} & 87.35 / 8.36 & 1.93 \\
  mlkem512 & sphincssha2128f & 2031.86 & 38334.88 & \textbf{51.76} / 49.66 & 3.33 \\
  hqc128 & mldsa44 & 2273.20 & 15597.85 & 71.99 / 28.91 & 3.02 \\
  hqc128 & falcon512 & 2286.71 & 11736.69 & 71.76 / 29.43 & 3.28 \\[0.5em]
  \multicolumn{6}{l}{\underline{\textit{[Legacy(KEM) + PQC(DS) Algorithms]}}} \\[0.25em]
  ffdhe2048 & mldsa44 & 1129.35 & 9498.89 & 86.35 / 11.88 & 2.11 \\
  ffdhe2048 & falcon512 & 1084.89 & 5504.62 & 84.53 / 12.80 & 1.90 \\
  ffdhe2048 & sphincssha2128f & 2099.34 & 37310.66 & \textbf{50.80} / 50.76 & 3.35 \\
  secp256r1 & mldsa44 & 1008.65 & 9129.04 & 87.34 / \textbf{8.05} & 1.82 \\
  x25519 & falcon512 & \textbf{989.28} & \textbf{5069.88} & 87.17 / 8.72 & 1.91 \\[0.5em]
  \multicolumn{6}{l}{\underline{\textit{[PQC(KEM) + Legacy(DS) Algorithms]}}} \\[0.25em]
  mlkem512 & rsa2048 & 1075.10 & 5161.51 & 85.06 / 11.65 & 1.97 \\
  mlkem512 & ecdsa-p256 & \textbf{1050.44} & 4587.73 & 89.40 / \textbf{6.67} & 1.87 \\
  mlkem512 & ed25519 & 1056.15 & \textbf{4518.04} & 89.91 / 6.70 & 2.00 \\
  hqc128 & rsa2048 & 2378.82 & 10241.93 & \textbf{70.11} / 30.91 & 3.08 \\
  hqc128 & ecdsa-p256 & 2319.31 & 9663.87 & 72.50 / 28.82 & 3.02 \\[0.5em]
  \bottomrule
  \end{tabular}
\end{table}

\begin{table}[htp!]
\footnotesize
  \caption{Multi-thread (10000 Concurrent Queries) DNS over TLS Benchmark Results by Algorithms (Security Level 1)}
  \label{tab:sl1_10000}
  \vspace{-0.5em}
  \begin{tabular}{llrrrr}
  \toprule
  \textbf{KEM} & \textbf{DS} & \textbf{Latency (ms)} & \textbf{Bandwidth (kB)} & \textbf{Client / Server CPU (\%)} & \textbf{Memory (\%)} \\
  \midrule
  \multicolumn{6}{l}{\underline{\textit{[Legacy(KEM) + Legacy(DS) Algorithms]}}} \\[0.25em]
  ffdhe2048 & rsa2048 & 12210.94 & 41409.08 & \textbf{85.04} / 16.81 & 3.24 \\
  ffdhe2048 & ecdsa-p256 & 11527.69 & 35695.47 & 88.77 / 12.26 & 3.11 \\
  ffdhe2048 & ed25519 & \textbf{11407.90} & \textbf{34987.19} & 88.88 / \textbf{11.83} & 3.2 \\
  secp256r1 & rsa2048 & 11477.12 & 37709.48 & 87.05 / 13.52 & 3.34 \\
  x25519 & rsa2048 & 11424.62 & 37058.06 & 87.45 / 13.28 & 3.28 \\[0.5em]
  \multicolumn{6}{l}{\underline{\textit{[PQC(KEM) + PQC(DS) Algorithms]}}} \\[0.25em]
  mlkem512 & mldsa44 & 11097.96 & 105253.99 & 92.55 / \textbf{8.02} & 3.16 \\
  mlkem512 & falcon512 & \textbf{10841.17} & \textbf{65355.13} & 91.48 / 9.18 & 3.21 \\
  mlkem512 & sphincssha2128f & 21392.90 & 383485.47 & \textbf{57.28} / 54.17 & 4.28 \\
  hqc128 & mldsa44 & 24288.86 & 156244.86 & 79.65 / 32.49 & 4.89 \\
  hqc128 & falcon512 & 24723.59 & 117645.91 & 75.73 / 31.36 & 4.55 \\[0.5em]
  \multicolumn{6}{l}{\underline{\textit{[Legacy(KEM) + PQC(DS) Algorithms]}}} \\[0.25em]
  ffdhe2048 & mldsa44 & 11867.66 & 95013.24 & 95.71 / 13.44 & 3.06 \\
  ffdhe2048 & falcon512 & 12014.42 & 55072.27 & 94.09 / 14.55 & 3.03 \\
  ffdhe2048 & sphincssha2128f & 22821.37 & 373292.91 & \textbf{56.34} / 55.74 & 5.13 \\
  secp256r1 & mldsa44 & \textbf{11062.00} & 91298.88 & 91.69 / \textbf{8.75} & 3.09 \\
  x25519 & falcon512 & 11249.60 & \textbf{50711.74} & 89.81 / 9.33 & 2.97 \\[0.5em]
  \multicolumn{6}{l}{\underline{\textit{[PQC(KEM) + Legacy(DS) Algorithms]}}} \\[0.25em]
  mlkem512 & rsa2048 & 11224.52 & 51639.30 & 87.52 / 12.28 & 3.06 \\
  mlkem512 & ecdsa-p256 & \textbf{11114.39} & 45836.77 & 92.10 / 7.08 & 3.01 \\
  mlkem512 & ed25519 & 11560.73 & \textbf{45110.46} & 94.02 / \textbf{7.05} & 3.47 \\
  hqc128 & rsa2048 & 24643.51 & 102661.15 & 73.80 / 31.28 & 4.69 \\
  hqc128 & ecdsa-p256 & 23974.78 & 96905.00 & \textbf{73.32} / 30.27 & 4.95 \\[0.5em]
  \bottomrule
  \end{tabular}
\end{table}

Compared to the 1,000-concurrent-query scenario, the 10,000-query benchmark as presented in the Table \ref{tab:sl1_10000} exposes sharper scalability limitations in both cryptographic processing and system resource usage. While high-performance combinations such as \texttt{MLKEM512 + Falcon512} and \texttt{MLKEM512 + MLDSA44} continue to demonstrate favorable latency-remaining near or slightly above 11,000 ms-client CPU usage now consistently exceeds 90-percent, suggesting that cryptographic operations are nearing saturation on the hardware. Algorithms with larger signature sizes and higher verification costs, notably \texttt{SPHINCS+} and \texttt{HQC}, experience significant performance degradation under this load: latencies surpass 24,000 ms, server CPU usage rises above 30-percent, and memory consumption increases to nearly 5-percent in extreme cases. Bandwidth overhead scales linearly with the query volume as expected, yet the total transfer size becomes critical-e.g., over 380 MB for \texttt{MLKEM512 + SPHINCS+}-raising deployment concerns in environments with limited throughput capacity. Hybrid combinations such as \texttt{FFDHE2048 + SPHINCS+} and \texttt{FFDHE2048 + Falcon512} exhibit similar symptoms, with both latency and CPU overhead widening their gap compared to mid-scale tests. These results affirm that while many PQC algorithms remain viable at moderate concurrency levels, certain schemes-especially those with hash-based signatures or large ciphertexts-present low performance in high-throughput use cases. 

\subsection{Analysis and Discussion of the Benchmarking Metrics}

The following analysis evaluates experimental benchmarking between PQC and conventional cryptographic techniques applied to DoT, DoH and DNSSEC configurations. The measurements were conducted in controlled Docker environments and focused on four primary performance metrics: Latency, bandwidth, CPU usage and memory consumption. The differences between the NIST security levels, the algorithm families (lattice, hash and code-based) and the protocol layers were investigated to understand the impact on real-world deployment.

\textit{Latency Usage:} Traditional algorithms such as FFDHE and RSA consistently exhibited faster handshake latency due to compact key sizes and minimal arithmetic complexity. Among PQC schemes, Falcon, ML-KEM, and ML-DSA demonstrated competitive latency, benefiting from efficient implementations and hardware-accelerated integer operations. In contrast, HQC and SPHINCS+ exhibited higher latency due to complex decoding logic or deep Merkle tree structures~\cite{NISTPQC}. The NIST security levels directly impacted latency. Higher levels require increased internal dimensions (e.g., lattice ring size, number of hash iterations), which translate to longer computation times during key encapsulation and signature verification. The latency was not only influenced by the bandwidth, but also by the computational effort. Algorithms with high arithmetic complexity or serialization overhead (e.g. syndrome decoding in HQC) significantly increased the handshake duration. Protocol differences (i.e., DoT vs. DoH) had a negligible effect on latency, with HTTP framing overhead being minimal.

\textit{Bandwidth Usage:} Traditional schemes transmitted smaller public keys and cipher-texts, keeping total handshake sizes compact. PQC algorithms, especially those based on hash and code constructs, required significantly more bandwidth due to large key and signature sizes. Hash-based algorithms (e.g., SPHINCS+) embed large hash trees into their signatures, leading to particularly high message overheads. Code-based schemes like HQC involve expanded cipher-texts for error correction. Lattice-based schemes (e.g., ML-KEM) offered a middle ground, achieving strong security with moderate bandwidth increases-due to their compact structured matrices and avoidance of large random data. DoH showed slightly higher bandwidth usage than DoT because of the extra overhead from HTTP/2 framing and headers. This difference is protocol-related and not caused by the cryptographic algorithms, including Legacy and PQC schemes.

\textit{CPU Usage:} CPU load during the establishment of the handshake exhibited a clear asymmetry between the client and the server. Across all algorithms, the client most likely experienced significantly higher CPU usage compared to the server. This reflects the dual burden of the client to perform both key encapsulation and signature verification, which are typically more computationally intensive than the decapsulation and signing operations of the server.  \textit{Lattice-based schemes} such as ML-KEM and ML-DSA, used vectorized modular arithmetic (e.g., AVX2) to accelerate operations \cite{zheng2024faster}. These schemes showed high CPU usage on the client side but relatively low load on the server, indicating efficient de-capsulation routines. \textit{Hash-based schemes} such as SPHINCS+, consumed fewer CPU cycles per clock but required many sequential operations during signature verification. This resulted in longer handshake durations and moderate CPU usage on both the client and the server. \textit{Code-based schemes} such as HQC, showed a more balanced CPU load between the client and server. The client’s CPU usage is lower because its operations, while involving some sequential processing, are less computationally intensive. In contrast, the server's higher CPU usage is driven by the error correction process, which involves complex and sequential syndrome decoding, causing a heavier load on the server side. \textit{Classical cryptographic schemes}, such as FFDHE with RSA or ECDSA, demonstrated stable and moderate CPU usage on both client and server sides. Their widespread optimization across platforms and decades of hardware support contributed to low cryptographic overheads, making them ideal for resource-constrained or latency-sensitive deployments. However, this efficiency comes at the cost of vulnerability to quantum attacks, underscoring the need to transition toward PQC alternatives despite the computational trade-offs. As security levels rise, the CPU load distribution between client and server becomes more evident. For higher security levels, the server CPU usage tends to increase due to the more complex de-capsulation and signature generation processes, while the client CPU usage generally decreases as key encapsulation and signature verification become more efficient, often benefiting from hardware acceleration or better optimizations. The underlying mathematical operations-such as number-theoretic transforms in lattice-based schemes, syndrome decoding in code-based schemes, or hash-tree traversal in hash-based schemes-significantly influence CPU usage. These operations affect not only the overall computational cost but also how efficiently the algorithms can be parallelized and optimized on modern hardware. Additionally, higher security levels typically increase computational demands, further impacting CPU utilization on both client and server systems.

\textit{Memory Usage:} Across all test scenarios, memory consumption remained within manageable limits. Traditional algorithms like RSA and ECDSA showed minimal memory overhead due to compact key sizes and optimized cryptographic routines. Among post-quantum algorithms, SPHINCS+ had the highest memory footprint, attributed to the deep hash tree structures used in its stateless signature generation. Conversely, ML-KEM and ML-DSA exhibited moderate memory usage due to their reliance on structured lattices and matrix arithmetic. HQC, while demanding more memory on the server during decoding, maintained acceptable RAM usage overall. None of the tested algorithms had significant impact on the allocated container memory, suggesting feasibility for DNS deployments even on modest hardware, provided CPU trade-offs are addressed.

\textit{Impact of DNSSEC:} DNSSEC introduced only minor increases in bandwidth when applied with traditional algorithms. When paired with PQC signatures (e.g., SPHINCS+), the increase became more pronounced, due to larger DNSKEY and RRSIG records, though it did not affect latency significantly. Applying DNSSEC without TLS caused no measurable latency penalty and introduced minimal resource overhead-highlighting its feasibility even in resource-constrained deployments. In combined DoT/DoH with DNSSEC scenarios, while bandwidth increased as expected, no significant differences were observed in other performance metrics such as latency or CPU usage between legacy and PQC algorithms.

\textit{Relationship of Metrics:}   Larger messages do contribute to increased bandwidth-but they are not the dominant cause of latency. Latency depends more heavily on algorithmic compute time and hardware optimization. High CPU usage may correlate with better performance, as it often indicates effective parallelism. For example, lattice-based schemes that fully utilized CPU resources achieved quicker handshakes despite high computational intensity.  Some schemes with low CPU usage still had poor latency due to serial or complex internal processes that failed to take advantage of available processing power. Bandwidth, CPU, and latency must be interpreted together rather than in isolation. No single metric accurately predicts overall performance across all protocols and algorithms.

\textbf{Multi-threaded Benchmarking Results}: Several metrics were evaluated for benchmarking purposes. The results of the analyses for the individual metrics are listed below:

\begin{itemize}[leftmargin=*]
    \item \textit{Latency:}  Compared to legacy algorithms, several PQC algorithm combinations- - especially ML-KEM with ML-DSA and Falcon- - showed strong performance at lower or comparable latency. This suggests that even under high concurrency, lattice-based methods can maintain responsiveness. However, algorithms such as SPHINCS+ and HQC required significantly more time to complete query transactions, highlighting their heavier cryptographic operations. Despite the transition to a multi-threaded environment, the overall latency trends remained consistent with those observed in the single-threaded test scenario.
    \item \textit{Bandwidth:} As expected, the PQC algorithms consume significantly more bandwidth than their legacy counterparts. This increase is due to the inherently larger key sizes and message structures of PQC cryptography. Furthermore, the behaviour was consistent with what was observed for single-threaded methods -the bandwidth values were about 100 times larger than a single-query session, reflecting the batch size of 100 concurrent queries per session.
    \item \textit{CPU Usage:}  CPU utilization on the client side remained consistently high for legacy algorithms. In contrast, PQC algorithms showed varying CPU demand depending on the underlying cryptographic family. Lattice-based schemes like ML-KEM and Falcon maintained higher client CPU and lower server CPU use, while hash-based (e.g., SPHINCS+) and code-based (e.g., HQC) algorithms presented more server CPU usage compared to Legacy and Lattice-based algorithms. These trends resembled the single-threaded results, with no significant deviation observed in the multi-threaded context.
    \item \textit{Memory Usage:}  Unlike CPU behavior, memory usage showed a noticeable increase due to multi-threaded execution. While legacy and PQC algorithms both experienced higher memory consumption, the increase was more pronounced in algorithms with inherently larger operational footprints-again, such as SPHINCS+ and HQC. This shift reflects the expected outcome of parallelized query execution placing additional demands on system memory.
\end{itemize}

\begin{figure}[htp!]
  \centering
  \begin{subfigure}[t]{0.49\linewidth}
 \includegraphics[width=\linewidth]{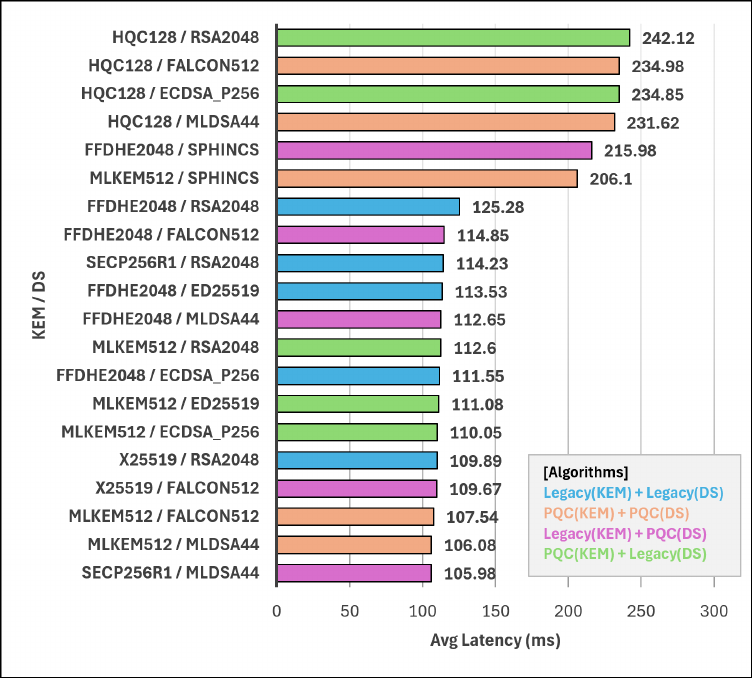}
    \Description{Bar chart showing Multi-Thread 100 queries PQC-DNS over TLS latency comparison under Security Level 1} 
  \end{subfigure}
  \hfill
  \begin{subfigure}[t]{0.49\linewidth}
    \includegraphics[width=\linewidth]{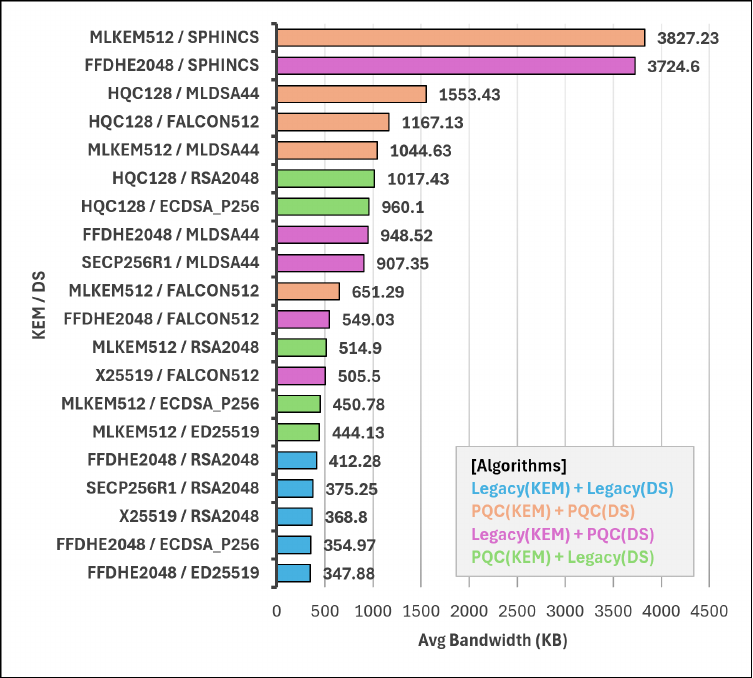}
    \Description{Bar chart showing Multi-Thread 100 queries PQC-DNS over TLS bandwidth comparison under Security Level 1} 
  \end{subfigure}
  \vspace{-1em}
  \caption{Multi-thread (100 Concurrent Queries) DoT Latency (left) and Bandwidth (right) comparison chart - Security Level 1}
  \label{fig:dot100}  
\end{figure}

\begin{figure}[ht]
  \centering
  \begin{subfigure}[t]{0.49\linewidth}
    \includegraphics[width=\linewidth]{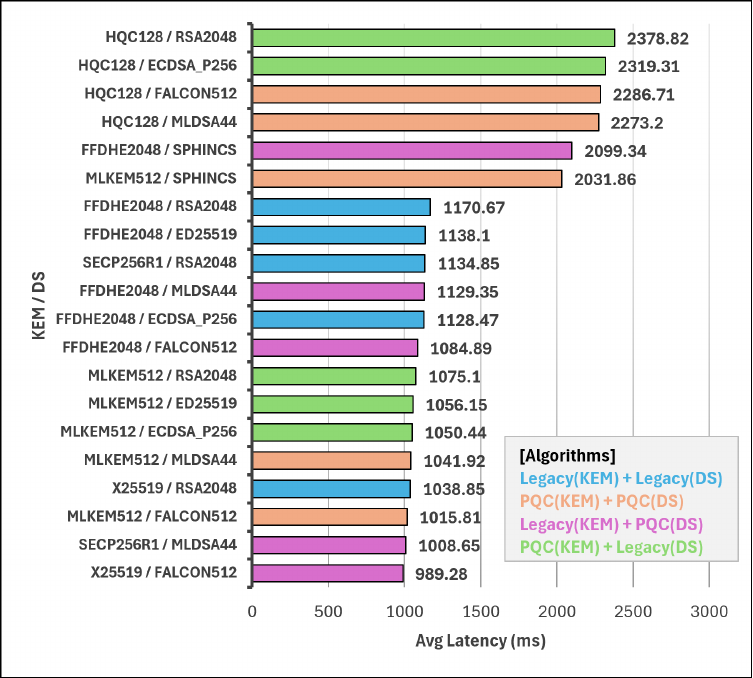}
    \Description{Bar chart showing Multi-Thread 1000 queries PQC-DNS over TLS latency comparison under Security Level 1} 
    \label{fig:multi_thread_dot_latency}
  \end{subfigure}
  \hfill
  \begin{subfigure}[t]{0.49\linewidth}
    \includegraphics[width=\linewidth]{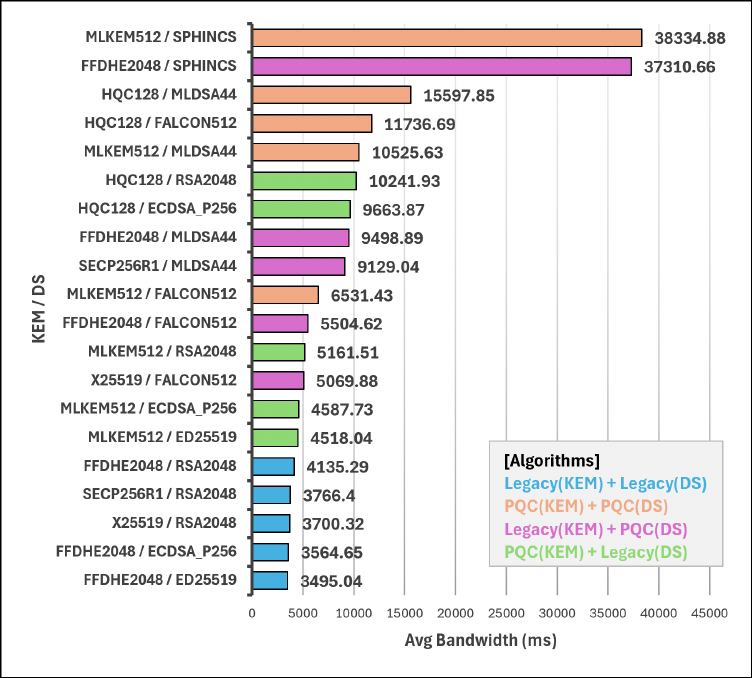}
    \Description{Bar chart showing Multi-Thread 1000 queries PQC-DNS over TLS bandwidth comparison under Security Level 1} 
    \label{fig:multi_thread_dot_bandwidth}
  \end{subfigure}
  \vspace{-1em}
  \caption{Multi-thread (1000 Concurrent Queries) DoT Latency (left) and Bandwidth (right) comparison chart - Security Level 1}
  \label{fig:dot1000}
\end{figure}

\begin{figure}[htp!]
  \centering
  \begin{subfigure}[t]{0.49\linewidth}
    \includegraphics[width=\linewidth]{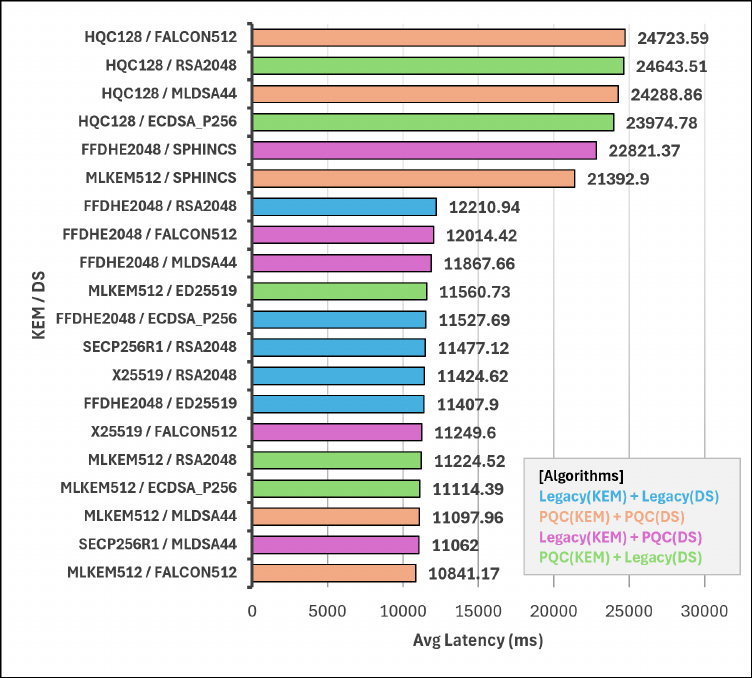}
    \Description{Bar chart showing Multi-Thread 10000 queries PQC-DNS over TLS latency comparison under Security Level 1} 
  \end{subfigure}
  \hfill
  \begin{subfigure}[t]{0.49\linewidth}
    \includegraphics[width=\linewidth]{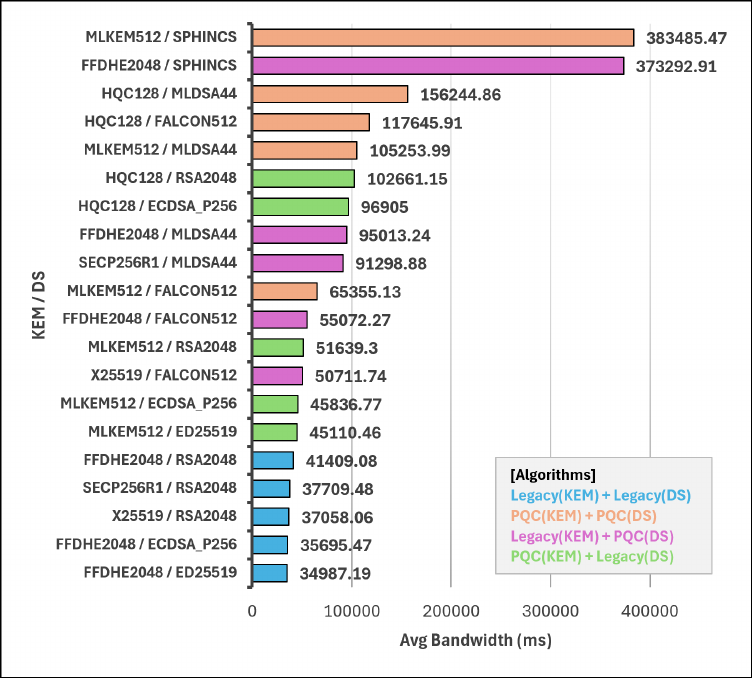}
    \Description{Bar chart showing Multi-Thread 10000 queries PQC-DNS over TLS bandwidth comparison under Security Level 1} 
  \end{subfigure}
  \vspace{-1em}
  \caption{Multi-thread (10000 Concurrent Queries) DoT Latency (left) and Bandwidth (right) comparison chart - Security Level 1}
  \label{fig:dot10000}
\end{figure}

Fig. \ref{fig:dot100}  presents multi-thread (i.e., 100 concurrent clients) DoT latency and bandwidth comparison chart based on Security Level 1 in an increasing order, Fig. \ref{fig:dot1000} shows for 1000 concurrent clients, and 10000 concurrent queries are displayed in Fig. \ref{fig:dot10000}. As with the single-threaded tests, DoH showed no significant performance differences compared to DoT, apart from the expected slight increase in bandwidth utilisation due to the HTTP header overhead. This pattern also remained consistent in the multi-threaded setting, indicating that protocol encapsulation does not result in additional performance degradation with concurrent query load.

\section{Limitations and Discussions}
\label{sec:discussion}

While our study presents a comprehensive performance evaluation of post-quantum DNS security mechanisms, several limitations and open issues remain that merit discussion.

\textit{Environment Configuration Complexity:} During the initial phase of this study, configuring a working post-quantum DNS environment posed considerable challenges. The integration of \texttt{OpenSSL}, \texttt{liboqs}, \texttt{oqsprovider}, and a forked \texttt{BIND} implementation (OQS-BIND) required substantial manual effort due to a lack of existing documentation or established deployment practices. Configuration mismatches, cryptographic interface inconsistencies, and versioning incompatibilities across these libraries significantly delayed experimental progress. This reflects a broader challenge in PQC deployment, which is the absence of mature tooling and interoperability support for production-ready DNS applications.

\textit{Hardware-Dependent Benchmark Results:} All performance metrics reported in this paper were obtained from a controlled local environment with fixed hardware specifications. While this ensures internal consistency and fair algorithmic comparisons, the results are inherently hardware-dependent and may not generalize to other platforms. In particular, CPU-bound operations such as lattice-based key encapsulations or hash-based signature verifications are sensitive to processor architecture, instruction set optimizations (e.g., AVX2), and system load. As such, these results should not be directly compared to benchmarks conducted in dissimilar environments. Moreover, network jitter, cross-traffic, and carrier-grade NAT behavior may impact real-world performance differently than reported here. Future work should extend this study to wide-area deployments or cloud-based DNS providers such as Cloudflare, Quad9, or Google DNS.

\textit{Security and Implementation Caveats:} Our security model assumes ideal implementations of PQC primitives. In practice, side-channel resistance is highly dependent on careful software and hardware engineering. Recent studies~\cite{hochstatter2023leaky} have shown that even constant-time implementations of Kyber and Dilithium can be vulnerable under real-world leakage. Furthermore, hybrid TLS deployments may expose endpoints to downgrade attacks if cipher suite negotiation is improperly enforced~\cite{ietf2023hybrid}. Ensuring downgrade resilience remains an open challenge, particularly during transitional deployments where backward compatibility with classical clients must be maintained.

\textit{DDoS and Operational Risks:} One critical operational concern is the susceptibility of PQC-enhanced handshakes to denial-of-service attacks. Recent work~\cite{shulman2023quantumddos} has demonstrated that the larger message sizes and increased CPU cycles required by post-quantum TLS can be exploited to launch resource exhaustion attacks against resolvers and servers. While our benchmarks included multi-threaded evaluations, future research must investigate rate-limiting, client puzzles, or adaptive throttling in high-load scenarios to maintain resolver availability.

\textit{Sustainability and Scalability Considerations:} The long-term viability of post-quantum DNS deployment depends not only on cryptographic soundness but also on sustainability and scalability across diverse Internet environments. Post-quantum algorithms-particularly lattice- and hash-based schemes-introduce nontrivial increases in computation time, memory footprint, and message size. These overheads raise concerns about energy efficiency, especially in edge and embedded deployments where power and thermal budgets are tightly constrained. For instance, the use of hash-based signatures like SPHINCS+ can amplify CPU cycles and bandwidth usage, potentially leading to increased data center energy demands and greater environmental impact. While lattice-based schemes such as MLKEM and Falcon offer more favorable performance-energy trade-offs, even these incur latency and CPU penalties compared to classical counterparts like X25519 or ECDSA. As global DNS infrastructure spans thousands of recursive and authoritative servers, even modest per-query increases can aggregate into substantial operational cost and carbon footprint. 
\textit{Scalability} is also challenged by the asymmetric nature of PQC performance, where server-side decapsulation and signature verification may dominate processing costs. In high-throughput scenarios, such as those faced by large DNS providers or CDNs, this imbalance could limit the capacity to handle peak query loads without architectural optimizations such as hardware acceleration, TLS session resumption, or offloading cryptographic functions to secure enclaves. To ensure scalable and sustainable deployment, future work must include energy profiling under realistic workloads, exploration of post-quantum hardware accelerators, and protocol-level optimizations such as lightweight hybrid modes and session caching. These enhancements are critical for aligning post-quantum security with operational feasibility and environmental responsibility.

\textit{Standardization and Deployment Barriers}:
Although NIST has finalized several PQC algorithms for standardization, many production-grade DNS resolvers (e.g., Unbound, PowerDNS) and TLS libraries still lack stable PQC integration. Our implementation is based on OQS-BIND and patched OpenSSL versions, which may diverge from mainstream stacks. Ensuring ecosystem-wide adoption will require coordinated efforts across DNS software vendors, TLS library maintainers, and browser manufacturers. Moreover, regulatory frameworks for DNS security (e.g., ICANN or DNSSEC root signing authorities) may need to update operational policies to accommodate hybrid or post-quantum-only deployments.

Despite these challenges, our results demonstrate that carefully selected PQC primitives-particularly MLKEM and Falcon-can enable post-quantum DNS security without incurring prohibitive performance penalties. Our findings can inform future standardization and deployment strategies for quantum-resilient DNS infrastructures.

\section{Conclusion and Future Work}
\label{sec:conclusion}

The findings presented in this study highlight the technical feasibility and practical considerations of deploying post-quantum cryptographic algorithms in DNS infrastructure, particularly in securing DNSSEC, DNS over TLS, DNS over HTTPS. By evaluating multiple algorithm classes across key performance metrics-latency, bandwidth, CPU, and memory usage-this work provides a foundation for informed decision-making in future DNS security designs. The demonstrated trade-offs emphasize that no single algorithm is universally optimal; rather, the suitability of a PQC scheme depends heavily on its operational context and intended role within the DNS protocol stack. As quantum-resistant standards continue to evolve, aligning DNS security mechanisms with these cryptographic advancements will be essential for maintaining long-term trust and resilience in global name resolution systems.


Future research should explore several key directions to advance the deployment of post-quantum secure DNS. First, evaluating the energy consumption and thermal behavior of PQC operations-particularly in edge computing environments-can help assess their sustainability and scalability. Second, integrating PQC mechanisms into emerging transport protocols such as DNS-over-QUIC (DoQ) would extend quantum resilience to modern, low-latency communication layers. 
Additionally, further investigation is needed into how PQC impacts resolver-side behaviors, including caching strategies and TTL (time-to-live) optimizations, which are critical for performance and efficiency at scale. Finally, conducting real-world user studies to measure latency perception and packet reliability under post-quantum DNS deployments will provide valuable insight into practical usability and quality-of-service implications.

Future work should also address these limitations by deploying PQC-enabled DNS in a production-like setting with real-world traffic patterns and multiple authoritative zones. In particular, scaling tests using distributed testbeds or cloud-based DNS resolvers could yield deeper insight into performance under load. 
Further, efforts should be made to measure protocol-level resilience, such as the effect of PQC on packet loss, retransmissions, and handshake failure rates in lossy or mobile networks. Lastly, given the fast-evolving nature of PQC standardization, continuous tracking of NIST and IETF progress will be essential to align future DNS deployments with approved cryptographic profiles.

\begin{acks}
This work is partly supported by the Spanish Ministry of Economy and Competitiveness (MINECO)-Program UNICO I+D under Grant TSI-063000-2021-54, Grant TSI-063000-2021-55, “ERDF A way of making Europe” project funded by MCIN/AEI/ 10.13039/501100011033  under grant PID2021-126431OB-I00 and Generalitat de Catalunya grant 2021 SGR 00770

\end{acks}

\bibliographystyle{ACM-Reference-Format}
\bibliography{References}


\begin{thebibliography}{43}


\ifx \showCODEN    \undefined \def \showCODEN     #1{\unskip}     \fi
\ifx \showDOI      \undefined \def \showDOI       #1{#1}\fi
\ifx \showISBNx    \undefined \def \showISBNx     #1{\unskip}     \fi
\ifx \showISBNxiii \undefined \def \showISBNxiii  #1{\unskip}     \fi
\ifx \showISSN     \undefined \def \showISSN      #1{\unskip}     \fi
\ifx \showLCCN     \undefined \def \showLCCN      #1{\unskip}     \fi
\ifx \shownote     \undefined \def \shownote      #1{#1}          \fi
\ifx \showarticletitle \undefined \def \showarticletitle #1{#1}   \fi
\ifx \showURL      \undefined \def \showURL       {\relax}        \fi
\providecommand\bibfield[2]{#2}
\providecommand\bibinfo[2]{#2}
\providecommand\natexlab[1]{#1}
\providecommand\showeprint[2][]{arXiv:#2}

\bibitem[Abirami and Naresh(2024)]%
        {10537516}
\bibfield{author}{\bibinfo{person}{S. Abirami} {and} \bibinfo{person}{R. Naresh}.} \bibinfo{year}{2024}\natexlab{}.
\newblock \showarticletitle{DNS Enhancement with DNSSEC and DoT for Enhanced Online Security}. In \bibinfo{booktitle}{\emph{2024 2nd International Conference on Networking and Communications (ICNWC)}}. \bibinfo{pages}{1--11}.
\newblock
\urldef\tempurl%
\url{https://doi.org/10.1109/ICNWC60771.2024.10537516}
\showDOI{\tempurl}


\bibitem[Ali and Chen({[n.\,d.]})]%
        {ali5230452titan}
\bibfield{author}{\bibinfo{person}{Basharat Ali} {and} \bibinfo{person}{Guihai Chen}.} \bibinfo{year}{[n.\,d.]}\natexlab{}.
\newblock \showarticletitle{Titan-Doh: Trust-Integrated Threat Adaptive Network for Post-Quantum Secure Dns Over Https}.
\newblock \bibinfo{journal}{\emph{Available at SSRN 5230452}} (\bibinfo{year}{[n.\,d.]}).
\newblock


\bibitem[Austein et~al\mbox{.}(2005)]%
        {RFC4033}
\bibfield{author}{\bibinfo{person}{Rob Austein}, \bibinfo{person}{Roy Arends}, \bibinfo{person}{Matt Larson}, \bibinfo{person}{Dan Massey}, {and} \bibinfo{person}{Scott Rose}.} \bibinfo{year}{2005}\natexlab{}.
\newblock \bibinfo{title}{DNS Security Introduction and Requirements}.
\newblock \bibinfo{howpublished}{RFC 4033}.
\newblock
\urldef\tempurl%
\url{https://tools.ietf.org/html/rfc4033}
\showURL{%
\tempurl}


\bibitem[Aydeger et~al\mbox{.}(2024)]%
        {aydeger2024towards}
\bibfield{author}{\bibinfo{person}{Abdullah Aydeger}, \bibinfo{person}{Engin Zeydan}, \bibinfo{person}{Awaneesh~Kumar Yadav}, \bibinfo{person}{Kasun~T Hemachandra}, {and} \bibinfo{person}{Madhusanka Liyanage}.} \bibinfo{year}{2024}\natexlab{}.
\newblock \showarticletitle{Towards a quantum-resilient future: Strategies for transitioning to post-quantum cryptography}. In \bibinfo{booktitle}{\emph{2024 15th International Conference on Network of the Future (NoF)}}. IEEE, \bibinfo{pages}{195--203}.
\newblock


\bibitem[Beernink(2022)]%
        {beernink2022taking}
\bibfield{author}{\bibinfo{person}{GJ Beernink}.} \bibinfo{year}{2022}\natexlab{}.
\newblock \emph{\bibinfo{title}{Taking the quantum leap: Preparing dnssec for post quantum cryptography}}.
\newblock \bibinfo{thesistype}{Master's\ thesis}. \bibinfo{school}{University of Twente}.
\newblock


\bibitem[Bozhko et~al\mbox{.}(2023)]%
        {bozhko2023performance}
\bibfield{author}{\bibinfo{person}{Jessica Bozhko}, \bibinfo{person}{Yacoub Hanna}, \bibinfo{person}{Ricardo Harrilal-Parchment}, \bibinfo{person}{Samet Tonyali}, {and} \bibinfo{person}{Kemal Akkaya}.} \bibinfo{year}{2023}\natexlab{}.
\newblock \showarticletitle{Performance evaluation of quantum-resistant TLS for consumer IoT devices}. In \bibinfo{booktitle}{\emph{2023 IEEE 20th Consumer Communications \& Networking Conference (CCNC)}}. IEEE, \bibinfo{pages}{230--235}.
\newblock


\bibitem[Chen et~al\mbox{.}(2016)]%
        {NISTIR8105}
\bibfield{author}{\bibinfo{person}{Lily Chen}, \bibinfo{person}{Stephen Jordan}, \bibinfo{person}{Yi-Kai Liu}, \bibinfo{person}{Dustin Moody}, {and} \bibinfo{person}{Rene Peralta}.} \bibinfo{year}{2016}\natexlab{}.
\newblock \bibinfo{booktitle}{\emph{Report on Post-Quantum Cryptography}}.
\newblock \bibinfo{type}{{T}echnical {R}eport} NISTIR 8105. \bibinfo{institution}{NIST}.
\newblock
\urldef\tempurl%
\url{https://doi.org/10.6028/NIST.IR.8105}
\showURL{%
\tempurl}


\bibitem[Chevalier et~al\mbox{.}(2022)]%
        {chevalier2022security}
\bibfield{author}{\bibinfo{person}{C{\'e}line Chevalier}, \bibinfo{person}{Ehsan Ebrahimi}, {and} \bibinfo{person}{Quoc-Huy Vu}.} \bibinfo{year}{2022}\natexlab{}.
\newblock \showarticletitle{On security notions for encryption in a quantum world}. In \bibinfo{booktitle}{\emph{International Conference on Cryptology in India}}. Springer, \bibinfo{pages}{592--613}.
\newblock


\bibitem[Cloudflare(2022)]%
        {CloudflarePQDNS}
\bibfield{author}{\bibinfo{person}{Cloudflare}.} \bibinfo{year}{2022}\natexlab{}.
\newblock \bibinfo{title}{Post-Quantum DNS and TLS}.
\newblock
\newblock
\urldef\tempurl%
\url{https://blog.cloudflare.com/post-quantum-for-dns/}
\showURL{%
\tempurl}


\bibitem[Dukhovni and Schwabe(2023)]%
        {dukhovni2023nonce}
\bibfield{author}{\bibinfo{person}{Viktor Dukhovni} {and} \bibinfo{person}{Peter Schwabe}.} \bibinfo{year}{2023}\natexlab{}.
\newblock \showarticletitle{Don't Reuse That Nonce: Failures in Randomness for Post-Quantum Signatures}.
\newblock \bibinfo{journal}{\emph{IACR Transactions on Cryptographic Hardware and Embedded Systems (TCHES)}} \bibinfo{volume}{2023}, \bibinfo{number}{1} (\bibinfo{year}{2023}), \bibinfo{pages}{1--24}.
\newblock
\urldef\tempurl%
\url{https://doi.org/10.46586/tches.v2023.i1.1-24}
\showDOI{\tempurl}


\bibitem[Goertzen and Stebila(2022)]%
        {goertzen2022arrf}
\bibfield{author}{\bibinfo{person}{Christopher Goertzen} {and} \bibinfo{person}{Douglas Stebila}.} \bibinfo{year}{2022}\natexlab{}.
\newblock \showarticletitle{ARRF: Application-layer Request-based Resource Fragmentation for Post-Quantum DNSSEC}. In \bibinfo{booktitle}{\emph{Proceedings of the 17th International Conference on Availability, Reliability and Security (ARES)}}.
\newblock
\urldef\tempurl%
\url{https://doi.org/10.1145/3538969.3544435}
\showDOI{\tempurl}


\bibitem[Goertzen(2023)]%
        {OQS-Bind}
\bibfield{author}{\bibinfo{person}{Jason Goertzen}.} \bibinfo{year}{2023}\natexlab{}.
\newblock \bibinfo{title}{OQS-BIND: PQC-enabled Bind9 using Open Quantum Safe's oqs-provider}.
\newblock
\newblock
\urldef\tempurl%
\url{https://github.com/Martyrshot/OQS-bind}
\showURL{%
\tempurl}


\bibitem[Goertzen and Stebila(2023)]%
        {goertzen2023post}
\bibfield{author}{\bibinfo{person}{Jason Goertzen} {and} \bibinfo{person}{Douglas Stebila}.} \bibinfo{year}{2023}\natexlab{}.
\newblock \showarticletitle{Post-quantum signatures in DNSSEC via request-based fragmentation}. In \bibinfo{booktitle}{\emph{International Conference on Post-Quantum Cryptography}}. Springer, \bibinfo{pages}{535--564}.
\newblock


\bibitem[Group(2023)]%
        {IETFPQDNSSEC}
\bibfield{author}{\bibinfo{person}{DNSOP~Working Group}.} \bibinfo{year}{2023}\natexlab{}.
\newblock \bibinfo{title}{Post-Quantum DNSSEC Considerations}.
\newblock \bibinfo{howpublished}{Internet Draft, IETF}.
\newblock
\urldef\tempurl%
\url{https://datatracker.ietf.org/doc/draft-dnsop-pqdnssig/}
\showURL{%
\tempurl}


\bibitem[Hochst{\"a}tter et~al\mbox{.}(2023)]%
        {hochstatter2023leaky}
\bibfield{author}{\bibinfo{person}{Jakob Hochst{\"a}tter}, \bibinfo{person}{Thomas Unterluggauer}, {and} \bibinfo{person}{Peter Schwabe}.} \bibinfo{year}{2023}\natexlab{}.
\newblock \showarticletitle{Leaky Kyber: Practical Side-Channel Attacks on Masked Kyber}. In \bibinfo{booktitle}{\emph{Proceedings of the 32nd USENIX Security Symposium}}. \bibinfo{publisher}{USENIX Association}.
\newblock
\urldef\tempurl%
\url{https://www.usenix.org/conference/usenixsecurity23/presentation/hochstatter}
\showURL{%
\tempurl}


\bibitem[Hoffman and McManus(2018)]%
        {RFC8484}
\bibfield{author}{\bibinfo{person}{Paul Hoffman} {and} \bibinfo{person}{Patrick McManus}.} \bibinfo{year}{2018}\natexlab{}.
\newblock \bibinfo{title}{DNS Queries over HTTPS (DoH)}.
\newblock \bibinfo{howpublished}{RFC 8484}.
\newblock
\urldef\tempurl%
\url{https://tools.ietf.org/html/rfc8484}
\showURL{%
\tempurl}


\bibitem[Hu et~al\mbox{.}(2016)]%
        {RFC7858}
\bibfield{author}{\bibinfo{person}{Zi Hu}, \bibinfo{person}{Liang Zhu}, \bibinfo{person}{John Heidemann}, \bibinfo{person}{Allison Mankin}, \bibinfo{person}{Duane Wessels}, {and} \bibinfo{person}{Paul Hoffman}.} \bibinfo{year}{2016}\natexlab{}.
\newblock \bibinfo{title}{Specification for DNS over Transport Layer Security (TLS)}.
\newblock \bibinfo{howpublished}{RFC 7858}.
\newblock
\urldef\tempurl%
\url{https://tools.ietf.org/html/rfc7858}
\showURL{%
\tempurl}


\bibitem[Hudaib and Hudaib(2014)]%
        {Hudaib2014DNS}
\bibfield{author}{\bibinfo{person}{Adam~Ali.Zare Hudaib} {and} \bibinfo{person}{Esra'a Ali~Zare Hudaib}.} \bibinfo{year}{2014}\natexlab{}.
\newblock \showarticletitle{DNS Advanced Attacks and Analysis}.
\newblock \bibinfo{journal}{\emph{International Journal of Computer Science and Security (IJCSS)}} \bibinfo{volume}{8}, \bibinfo{number}{2} (\bibinfo{date}{April} \bibinfo{year}{2014}), \bibinfo{pages}{63--74}.
\newblock
\urldef\tempurl%
\url{https://www.cscjournals.org/library/manuscriptinfo.php?mc=IJCSS-905}
\showURL{%
\tempurl}


\bibitem[Hülsing et~al\mbox{.}(2021)]%
        {Huelsing2021PQTLS}
\bibfield{author}{\bibinfo{person}{Andreas Hülsing}, \bibinfo{person}{Joost Rijneveld}, {and} \bibinfo{person}{Douglas Stebila}.} \bibinfo{year}{2021}\natexlab{}.
\newblock \showarticletitle{PQC in TLS: How to Make and Break It}. In \bibinfo{booktitle}{\emph{30th USENIX Security Symposium}}.
\newblock
\urldef\tempurl%
\url{https://www.usenix.org/conference/usenixsecurity21/presentation/hulsing}
\showURL{%
\tempurl}


\bibitem[Jafarli(2022)]%
        {jafarli2022providing}
\bibfield{author}{\bibinfo{person}{Sevinj Jafarli}.} \bibinfo{year}{2022}\natexlab{}.
\newblock \emph{\bibinfo{title}{Providing DNS Security in Post-Quantum Era with Hash-Based Signatures}}.
\newblock \bibinfo{thesistype}{Master's\ thesis}. \bibinfo{school}{University of Twente}.
\newblock


\bibitem[Labs(2023)]%
        {NLnetLabsPQC}
\bibfield{author}{\bibinfo{person}{NLnet Labs}.} \bibinfo{year}{2023}\natexlab{}.
\newblock \bibinfo{title}{PQC Readiness in DNS}.
\newblock
\newblock
\urldef\tempurl%
\url{https://nlnetlabs.nl}
\showURL{%
\tempurl}


\bibitem[McGowan et~al\mbox{.}(2025)]%
        {mcgowan2025security}
\bibfield{author}{\bibinfo{person}{Cameron McGowan}, \bibinfo{person}{James Liu}, {and} \bibinfo{person}{Sushmita Ruj}.} \bibinfo{year}{2025}\natexlab{}.
\newblock \showarticletitle{Security Considerations for Post-Quantum Signatures in DNSSEC via Request-Based Fragmentation}. In \bibinfo{booktitle}{\emph{Companion Proceedings of the ACM on Web Conference 2025}}. \bibinfo{pages}{1189--1193}.
\newblock


\bibitem[{National Institute of Standards and Technology}(2022)]%
        {NISTPQC}
\bibfield{author}{\bibinfo{person}{{National Institute of Standards and Technology}}.} \bibinfo{year}{2022}\natexlab{}.
\newblock \bibinfo{title}{Post-Quantum Cryptography Standardization}.
\newblock \bibinfo{howpublished}{\url{https://csrc.nist.gov/projects/post-quantum-cryptography}}.
\newblock


\bibitem[{National Institute of Standards and Technology (NIST)}(2015)]%
        {nist:sp800-152}
\bibfield{author}{\bibinfo{person}{{National Institute of Standards and Technology (NIST)}}.} \bibinfo{year}{2015}\natexlab{}.
\newblock \bibinfo{booktitle}{\emph{{SP 800-152: A Profile for U.S. Federal Cryptographic Key Management Systems}}}.
\newblock \bibinfo{type}{Special Publication} SP 800-152. \bibinfo{institution}{NIST Computer Security Division}.
\newblock
\urldef\tempurl%
\url{https://nvlpubs.nist.gov/nistpubs/SpecialPublications/NIST.SP.800-152.pdf}
\showURL{%
\tempurl}
\newblock
\shownote{Final publication, October 30, 2015}.


\bibitem[Pan et~al\mbox{.}(2024)]%
        {pan2024double}
\bibfield{author}{\bibinfo{person}{Syed W~Shah Pan}, \bibinfo{person}{Din Duc~Nha Nguyen}, \bibinfo{person}{Robin Doss}, \bibinfo{person}{Warren Armstrong}, \bibinfo{person}{Praveen Gauravaram}, {et~al\mbox{.}}} \bibinfo{year}{2024}\natexlab{}.
\newblock \showarticletitle{Double-Signed Fragmented DNSSEC for Countering Quantum Threat}.
\newblock \bibinfo{journal}{\emph{arXiv preprint arXiv:2411.07535}} (\bibinfo{year}{2024}).
\newblock


\bibitem[Project(2022)]%
        {PQDNS2022}
\bibfield{author}{\bibinfo{person}{PQDNS Project}.} \bibinfo{year}{2022}\natexlab{}.
\newblock \bibinfo{title}{Post-Quantum Secure DNS Prototype}.
\newblock
\newblock
\urldef\tempurl%
\url{https://pqdns.dev}
\showURL{%
\tempurl}


\bibitem[Raavi et~al\mbox{.}(2024)]%
        {raavi2024securing}
\bibfield{author}{\bibinfo{person}{Manohar Raavi}, \bibinfo{person}{Simeon Wuthier}, {and} \bibinfo{person}{Sang-Yoon Chang}.} \bibinfo{year}{2024}\natexlab{}.
\newblock \showarticletitle{Securing Post-Quantum DNSSEC Against Fragmentation Mis-Association Threat}. In \bibinfo{booktitle}{\emph{ICC 2024-IEEE International Conference on Communications}}. IEEE, \bibinfo{pages}{97--102}.
\newblock


\bibitem[Rawat and Jhanwar(2023a)]%
        {rawat2023qbf}
\bibfield{author}{\bibinfo{person}{Arpit Rawat} {and} \bibinfo{person}{Mayank Jhanwar}.} \bibinfo{year}{2023}\natexlab{a}.
\newblock \showarticletitle{QBF: QNAME-Based Fragmentation for Post-Quantum DNSSEC}. In \bibinfo{booktitle}{\emph{IEEE Symposium on Security and Privacy Workshops (SPW)}}.
\newblock
\urldef\tempurl%
\url{https://doi.org/10.1109/SPW59501.2023.00053}
\showDOI{\tempurl}


\bibitem[Rawat and Jhanwar(2023b)]%
        {rawat2023sldnssec}
\bibfield{author}{\bibinfo{person}{Arpit Rawat} {and} \bibinfo{person}{Mayank Jhanwar}.} \bibinfo{year}{2023}\natexlab{b}.
\newblock \showarticletitle{SL-DNSSEC: A Size-Optimized Post-Quantum Secure DNSSEC Protocol Using KEM and MAC}. In \bibinfo{booktitle}{\emph{2023 ACM Asia Conference on Computer and Communications Security (AsiaCCS)}}.
\newblock
\urldef\tempurl%
\url{https://doi.org/10.1145/3579856.3592816}
\showDOI{\tempurl}


\bibitem[Rawat and Jhanwar(2023c)]%
        {rawat2023turbodns}
\bibfield{author}{\bibinfo{person}{Arpit Rawat} {and} \bibinfo{person}{Mayank Jhanwar}.} \bibinfo{year}{2023}\natexlab{c}.
\newblock \showarticletitle{TurboDNS: Efficient Post-Quantum Secure DNSSEC over TCP}. In \bibinfo{booktitle}{\emph{2023 IEEE International Conference on Distributed Computing Systems (ICDCS)}}.
\newblock
\urldef\tempurl%
\url{https://doi.org/10.1109/ICDCS58692.2023.00123}
\showDOI{\tempurl}


\bibitem[Rawat and Jhanwar(2023d)]%
        {rawat2023post}
\bibfield{author}{\bibinfo{person}{Aditya~Singh Rawat} {and} \bibinfo{person}{Mahabir~Prasad Jhanwar}.} \bibinfo{year}{2023}\natexlab{d}.
\newblock \showarticletitle{Post-quantum DNSSEC over UDP via QNAME-Based Fragmentation}. In \bibinfo{booktitle}{\emph{International Conference on Security, Privacy, and Applied Cryptography Engineering}}. Springer, \bibinfo{pages}{66--85}.
\newblock


\bibitem[Rawat and Jhanwar(2024a)]%
        {rawat2024post}
\bibfield{author}{\bibinfo{person}{Aditya~Singh Rawat} {and} \bibinfo{person}{Mahabir~Prasad Jhanwar}.} \bibinfo{year}{2024}\natexlab{a}.
\newblock \showarticletitle{Post-Quantum DNSSEC with Faster TCP Fallbacks}. In \bibinfo{booktitle}{\emph{International Conference on Cryptology in India}}. Springer, \bibinfo{pages}{212--236}.
\newblock


\bibitem[Rawat and Jhanwar(2024b)]%
        {rawat2024quantum}
\bibfield{author}{\bibinfo{person}{Aditya~Singh Rawat} {and} \bibinfo{person}{Mahabir~Prasad Jhanwar}.} \bibinfo{year}{2024}\natexlab{b}.
\newblock \showarticletitle{Quantum-safe Signatureless DNSSEC}.
\newblock \bibinfo{journal}{\emph{Cryptology ePrint Archive}} (\bibinfo{year}{2024}).
\newblock


\bibitem[Schutijser et~al\mbox{.}(2024)]%
        {schutijser2024testbed}
\bibfield{author}{\bibinfo{person}{Caspar Schutijser}, \bibinfo{person}{Elmer~EH Lastdrager}, \bibinfo{person}{Ralph Koning}, {and} \bibinfo{person}{Cristian~EW Hesselman}.} \bibinfo{year}{2024}\natexlab{}.
\newblock \showarticletitle{A testbed to evaluate quantum-safe cryptography in DNSSEC}. In \bibinfo{booktitle}{\emph{DNS and Internet Naming Research Directions, DINR 2024}}.
\newblock


\bibitem[Sengupta et~al\mbox{.}(2024)]%
        {10486930}
\bibfield{author}{\bibinfo{person}{Jayasree Sengupta}, \bibinfo{person}{Mike Kosek}, \bibinfo{person}{Justus Fries}, \bibinfo{person}{Simone Ferlin-Reiter}, {and} \bibinfo{person}{Vaibhav Bajpai}.} \bibinfo{year}{2024}\natexlab{}.
\newblock \showarticletitle{On Cross-Layer Interactions of QUIC, Encrypted DNS and HTTP/3: Design, Evaluation, and Dataset}.
\newblock \bibinfo{journal}{\emph{IEEE Transactions on Network and Service Management}} \bibinfo{volume}{21}, \bibinfo{number}{3} (\bibinfo{year}{2024}), \bibinfo{pages}{2992--3007}.
\newblock
\urldef\tempurl%
\url{https://doi.org/10.1109/TNSM.2024.3383787}
\showDOI{\tempurl}


\bibitem[Shulman and Waidner(2023)]%
        {shulman2023quantumddos}
\bibfield{author}{\bibinfo{person}{Haya Shulman} {and} \bibinfo{person}{Michael Waidner}.} \bibinfo{year}{2023}\natexlab{}.
\newblock \showarticletitle{Post-Quantum DDoS: How Quantum-Safe Handshakes Can Be Weaponized}. In \bibinfo{booktitle}{\emph{Proceedings of the Network and Distributed System Security Symposium (NDSS)}}.
\newblock
\urldef\tempurl%
\url{https://www.ndss-symposium.org/ndss-paper/post-quantum-ddos-how-quantum-safe-handshakes-can-be-weaponized/}
\showURL{%
\tempurl}


\bibitem[Stebila et~al\mbox{.}(2016)]%
        {Stebila2016Hybrid}
\bibfield{author}{\bibinfo{person}{Douglas Stebila}, \bibinfo{person}{Scott Fluhrer}, {and} \bibinfo{person}{Shay Gueron}.} \bibinfo{year}{2016}\natexlab{}.
\newblock \showarticletitle{Hybrid key exchange in TLS 1.3}.
\newblock \bibinfo{journal}{\emph{IACR Cryptology ePrint Archive}}  \bibinfo{volume}{2016} (\bibinfo{year}{2016}), \bibinfo{pages}{1008}.
\newblock
\urldef\tempurl%
\url{https://eprint.iacr.org/2016/1008}
\showURL{%
\tempurl}


\bibitem[Stebila et~al\mbox{.}(2023)]%
        {ietf2023hybrid}
\bibfield{author}{\bibinfo{person}{Douglas Stebila}, \bibinfo{person}{Scott Fluhrer}, {and} \bibinfo{person}{Martin Thomson}.} \bibinfo{year}{2023}\natexlab{}.
\newblock \bibinfo{title}{Hybrid Key Exchange in TLS 1.3}.
\newblock \bibinfo{howpublished}{Internet-Draft draft-ietf-tls-hybrid-design-05, IETF}.
\newblock
\urldef\tempurl%
\url{https://datatracker.ietf.org/doc/html/draft-ietf-tls-hybrid-design-05}
\showURL{%
\tempurl}
\newblock
\shownote{Work in Progress}.


\bibitem[Team(2023)]%
        {GooglePQ2023}
\bibfield{author}{\bibinfo{person}{Google~Security Team}.} \bibinfo{year}{2023}\natexlab{}.
\newblock \bibinfo{title}{Experimenting with post-quantum TLS in Chrome}.
\newblock
\newblock
\urldef\tempurl%
\url{https://security.googleblog.com/2023/08/experimenting-with-post-quantum-tls-in.html}
\showURL{%
\tempurl}


\bibitem[Wood et~al\mbox{.}(2023)]%
        {ietf2023interop}
\bibfield{author}{\bibinfo{person}{Chris Wood}, \bibinfo{person}{Martin Thomson}, {and} \bibinfo{person}{Daniel Migault}.} \bibinfo{year}{2023}\natexlab{}.
\newblock \bibinfo{title}{Interoperability Considerations for Post-Quantum TLS}.
\newblock \bibinfo{howpublished}{Internet-Draft draft-wood-tls-pqc-interoperability-01, IETF}.
\newblock
\urldef\tempurl%
\url{https://datatracker.ietf.org/doc/html/draft-wood-tls-pqc-interoperability-01}
\showURL{%
\tempurl}
\newblock
\shownote{Work in Progress}.


\bibitem[Xagawa(2024)]%
        {xagawa2024signatures}
\bibfield{author}{\bibinfo{person}{Keita Xagawa}.} \bibinfo{year}{2024}\natexlab{}.
\newblock \showarticletitle{Signatures with memory-tight security in the quantum random oracle model}. In \bibinfo{booktitle}{\emph{Annual International Conference on the Theory and Applications of Cryptographic Techniques}}. Springer, \bibinfo{pages}{30--58}.
\newblock


\bibitem[Zhang et~al\mbox{.}(2022)]%
        {Zhang2022SPHINCS}
\bibfield{author}{\bibinfo{person}{Yuan Zhang}, \bibinfo{person}{Zi Hu}, {and} \bibinfo{person}{Kevin Schomp}.} \bibinfo{year}{2022}\natexlab{}.
\newblock \showarticletitle{Evaluating SPHINCS+ for DNSSEC}. In \bibinfo{booktitle}{\emph{Proceedings of the NDSS Workshop on DNS Privacy}}.
\newblock


\bibitem[Zheng et~al\mbox{.}(2024)]%
        {zheng2024faster}
\bibfield{author}{\bibinfo{person}{Jieyu Zheng}, \bibinfo{person}{Haoliang Zhu}, \bibinfo{person}{Yifan Dong}, \bibinfo{person}{Zhenyu Song}, \bibinfo{person}{Zhenhao Zhang}, \bibinfo{person}{Yafang Yang}, {and} \bibinfo{person}{Yunlei Zhao}.} \bibinfo{year}{2024}\natexlab{}.
\newblock \showarticletitle{Faster Post-quantum TLS 1.3 Based on ML-KEM: Implementation and Assessment}. In \bibinfo{booktitle}{\emph{European Symposium on Research in Computer Security}}. Springer, \bibinfo{pages}{123--143}.
\newblock


\end{thebibliography}



\end{document}